\documentclass[%
 reprint,
 amsmath,amssymb,amsthm,
 nofootinbib,
 aps,
]{revtex4-1}

\usepackage{dcolumn}
\usepackage{bm}
\usepackage[mathlines]{lineno}
\usepackage{graphicx}
\usepackage[font=small,labelfont=bf]{caption}
\usepackage{dirtytalk}
\usepackage{csquotes}

\newtheorem{theorem}{Theorem}

\begin{document}
\preprint{APS/123-QED}

\title{Quantum Mechanics from Relational Properties \\Part I: Basic Formulation}

\author{Jianhao M. Yang}
\email{jianhao.yang@alumni.utoronto.ca}
\affiliation{San Diego, CA 92121, USA
}

\date{\today}

\begin{abstract}
Non-relativistic quantum mechanics is reformulated here based on the idea that relational properties among quantum systems, instead of the independent properties of a quantum system, are the most fundamental elements to construct quantum mechanics. This idea, combining with the emphasis that measurement of a quantum system is a bidirectional interaction process, leads to a new framework to calculate the probability of an outcome when measuring a quantum system. In this framework, the most basic variable is the relational probability amplitude. Probability is calculated as summation of weights from the potential alternative measurement configurations. The properties of quantum systems, such as superposition and entanglement, are manifested through the rules of counting the alternatives. Wave function and reduced density matrix are derived from the relational probability amplitude matrix. They are found to be secondary mathematical tools that equivalently describe a quantum system without explicitly calling out the measuring system. Schr\"{o}dinger Equation is obtained when there is no entanglement in the relational probability amplitude matrix. Feynman Path Integral is used to calculate the relational probability amplitude, and is further generalized to formulate the reduced density matrix. In essence, quantum mechanics is reformulated as a theory that describes physical systems in terms of relational properties.
\begin{description}
\item[Keywords] Relational Quantum Mechanics, Measurement Probability, Summation of Alternatives, \\Entanglement, Quantum Reference Frame
\end{description}
\end{abstract}
\maketitle

\section{Introduction}
\label{intro}
Although quantum mechanics is one of the most successful physical theories and has been experimentally confirmed extensively, there are many fundamental questions still left unanswered. For instance, the origin of probability in quantum mechanics is not clearly understood. It is still a curiosity why the probability is calculated as the absolute square of a complex number. The meaning of wave function, especially the interpretation of wave function collapse in a measurement, has been always a debated topic. These questions were not fully addressed by the traditional Copenhagen Interpretation~\cite{Bohr}. Over the years in the modern history of quantum physics, many more theories and interpretations have been developed. These include the many-worlds interpretation~\cite{Everett, Wheeler57, DeWitt70}, consistent histories~\cite{Griff84, Griff96, Omnes, GMH90}, decoherent theory~\cite{Zurek82, Zurek03, Schlosshauer04}, relational interpretations~\cite{Rovelli96, Rovelli07}, quantum Bayesian theory~\cite{Fuchs02, Fuchs13}, and many others. Along the development of these interpretations, one noticeable idea is the realization that a quantum state is relative in nature. That is, an observer independent quantum state is not necessarily the basic description of a quantum system. In the early days of quantum mechanics, Bohr had already emphasized that the description of a quantum system depends on the measuring apparatus~\cite{Bohr35, Jammer74}. Ref.~\cite{Everett} recognized that a quantum state of a subsystem is only meaningful relative to a given state of the rest of the system. Similarly, in developing the theory of decoherence induced by environment, Ref.~\cite{Zurek82} concluded that correlation information between two quantum systems is more basic than the properties of the quantum systems themselves. Relational Quantum Mechanics (RQM) has pursued this idea to the furthest extend. RQM is inspired by the basic principle from Einstein's Special Relativity. In the context of RQM, a quantum system should be described relative to another system, there is no absolute state for a quantum system. Specifically, the main idea of RQM is stated as following,
\begin{displayquote}
\textit{Quantum mechanics is a theory about the physical description of physical system relative to other systems, and this is a complete description of the world~\cite{Rovelli96}.}
\end{displayquote}
This statement appears radical but reflects the fact that quantum mechanics was originally developed as a theory to describe the experimental observations of a quantum system in a measurement. When we state that the observing system records the measurement results of a variable of the observed system, it means that a correlation between the observed system and the observing system is established through physical interaction. By reading the pointer variable in the observing system, one can infer the value of variable in the observed system. In this sense, quantum theory does not describe the independent properties of a quantum system. Instead, it describes the relation among quantum systems, and how correlation is established through physical interaction during measurement. The reality of a quantum system is only meaningful in the context of measurement by another system.

The idea of RQM is thought provoking. It essentially implies two aspects of relativity. The first aspect of RQM is to insist that a quantum system must be described relative to a reference system. The reference system is arbitrarily selected. It can be an apparatus in a measurement setup, or another system in the environment. A quantum system can be described differently relative to different reference systems. The reference system itself is also a quantum system, which is called a quantum reference frame (QRF). There are extensive research activities on QRF, particularly how to ensure consistent descriptions when switching QRFs~\cite{QRF1, QRF2, QRF3, QRF4, QRF5, QRF6, QRF7, QRF8, QRF9, QRF10, QRF11, QRF12, QRF13, QRF14, QRF15, QRF16, Brukner, Hoehn2018}. Noticeably, Ref.~\cite{Brukner, Hoehn2018, Hoehn2019, Yang2020, Ballesteros} completely abandon any external reference system and the concept of absolute state. Physical description is constructed using relational variables from the very beginning within the framework of traditional quantum mechanics. In addition, all reference systems are treated as quantum systems instead of some kinds of abstract entities. Treating a reference frame as a classical system, such as how the relativity theory does, should be considered as an approximation of a more fundamental theory that is based on QRF.

The second aspect of RQM is more fundamental. Since the relational properties between two quantum systems are considered more basic than the independent properties of one system, the relational properties, instead of the independent properties, of quantum systems should be considered as a starting point for constructing the formulation of quantum mechanics itself. Questions associated with this aspect of RQM include how to quantify the relational properties between two quantum systems, and how to reconstruct a quantum mechanics theory from relational properties. Note that the relational properties themselves are relative to a QRF. Different observers can ascribe a quantum system with different sets of relational properties relative to their choices of QRFs. 

It is this second aspect of RQM that inspires our works here. Traditional quantum mechanics always starts with an observer-independent quantum state. It is of interest to see if a quantum theory constructed based on relational properties can address some of the unanswered fundamental questions mentioned earlier. Such reconstruction program was initiated in Ref.~\cite{Rovelli96} and had some successes, for example, in deriving the Schr\"{o}dinger Equation. This reconstruction is based on quantum logic approach. Alternative reconstruction that follows the RQM principle but based on information theory is also developed~\cite{Hoehn2014, Hoehn2015}. These reconstructions appear rather abstract, not closely connect to the physical process of a quantum measurement. We believe that the relational properties should be identified in a measurement event given the idea that the reality of a quantum system is only meaningful in the context of measurement by another system.

The goal of this paper is to continue the program of reconstructing the formulation of quantum mechanics with the starting point that the relational properties are the most basic elements. What is novel in our approach is a new framework for calculating the probability of an outcome when measuring a quantum system. Such a framework is fundamental in deriving basic laws of quantum mechanics, so we briefly describe it here. In searching for the appropriate relational properties as the starting elements for the reconstruction, we recognize that a physical measurement is a probe-response interaction process between the measured system and the measuring apparatus. This important aspect of measurement process seems being overlooked in other reconstruction efforts. Our framework for calculating the probability, on the other hand, explicitly models this bidirectional process. As such, the probability can be derived from product of two quantities and each quantity is associated with a unidirectional process. We call such quantity relational probability amplitude. When two quantum systems interact, there are many alternative configurations for such two-way process. Each alternative is assigned with a weight that is a product of two relational probability amplitudes associated with the configuration. The probability of a measurement outcome is then postulated to be proportional to the summation of such weights from all the applicable configurations. Thus, the task of calculating the probability is reduced to counting the applicable alternatives. The properties of the measured system are manifested through the rules to count the alternatives. Another aspect of novelty of this framework is the introduction of the concept of entanglement to the relational properties. When the quantum system is entangled with the apparatus, the rule of counting the alternatives is adjusted accordingly due to the availability of inference information. Furthermore, the entanglement measure quantifies the difference between time evolution and quantum measurement. Lastly, we show that such framework to calculate probability amplitude can be explicitly implemented using Feynman Path Integral~\cite{Feynman48}.

The impacts of this framework are fundamental as it is the basis for deriving the formulations that are mathematically equivalent to the laws in traditional quantum mechanics. The formulation for calculating the probability of finding the system in an eigenstate is equivalent to Born's rule, but with a new insight: the fact it is an absolute square of a complex number is a consequence that a quantum measurement is a bidirectional process. Wave function is found to be a mathematical tool representing the summation of relational probability amplitude. Thus, the notion of wave function collapse during measurement is just a consequence of changes of relational properties. Schr\"{o}dinger equation can be derived when the entanglement measure between the observed quantum system and the observing system is zero and unchanged. On the other hand, when there is change in the entanglement measure, quantum measurement theory is obtained. 

Although the formulation presented here is mathematically equivalent to the traditional quantum mechanics, the theory presented here provides new understanding on the origin of quantum probability. It shows that an essence of quantum mechanics is a new set of rules to calculate the measurement probability from an interaction process. The most important outcome of this paper is that quantum mechanics can be constructed with the relational properties among quantum systems as the most fundamental building blocks.

The paper is organized as following. We first clarify the definitions of key terminologies in Section \ref{sec:concept}. Section \ref{sec:results} gives the main results of the paper. The postulates and frameworks to calculate quantum probability is provided in Section \ref{subsec:postulates} and \ref{subsec:WF}. In Section \ref{subsec:TV} to \ref{subsec:generalTV} formulation for time evolution of a quantum system is developed, and the conditions when Schr\"{o}dinger equation can be recovered are analyzed. In Section \ref{sec:conclusion}, we provide a comparison between this works and the original RQM theory, discuss the limitations, and summarize the conclusions. An explicit calculation of the relational probability amplitude using Feynman Path Integral formulation is presented in Section \ref{sec:PI}.

\section{Definitions of Key Terminologies}
\label{sec:concept}

\subsection{Quantum System, Apparatus, and Observer}
\label{subsec:definition}
To avoid potential confusion, it is useful to define several key terms before proceeding. A \textit{Quantum System}, denoted by symbol $S$, is an object under study and follows the postulates that will be introduced in next section. An \textit{Apparatus}, denoted as $A$, can refer to the measuring devices, the environment that $S$ is interacting with, or the system from which $S$ is created. It is another quantum system that can interact with $S$, can acquire or encode information from $S$. We will strictly follow the assumptions that all systems are quantum systems, including any apparatus. Depending on the selection of observer, the boundary between a system and an apparatus can change. For example, in a measurement setup, the measuring system is an apparatus $A$, the measured system is $S$. However, the composite system $S+A$ as a whole can be considered a single system, relative to another apparatus $A'$. In an ideal measurement to measure an observable of $S$, the apparatus is designed in such a way that at the end of the measurement, the pointer state of $A$ has a distinguishable, one to one correlation with the eigenvalue of the observable of $S$.

The definition of \textit{Observer} is associated with an apparatus. An observer, denoted as $\cal{O}$, is an intelligent entity who can operate and read the pointer variable of the apparatus. This can be a human being, or an artificial intelligent computer. The distinction between an observer and an apparatus is that an apparatus directly interacts with $S$, while an observer does not. Whether or not this observer is a quantum system is irrelevant in our formulation. However, there is a restriction that is imposed by the principle of locality. An observer is defined to be physically local to the apparatus he associates with. This prevents the situation that $\cal{O}$ can instantaneously read the pointer variable of the apparatus that is space-like separated from $\cal{O}$. Receiving the information from $A$ at a speed faster than the speed of light is prohibited. This locality requirement is crucial in resolving the EPR argument~\cite{EPR, Rovelli07}. An observer cannot be associated with two or more apparatuses in the same time if these apparatuses are space-like separated. 

A \textit{Quantum Reference Frame (QRF)} is a quantum system where all the descriptions of the relational properties between $S$ and $A$ is referred to. There can be multiple QRFs. How the descriptions are transformed when switching QRFs is not in the scope of this study. But we expect the theories developed in Ref.~\cite{Brukner, Hoehn2018, Yang2020} can be applicable here. In this paper, we only consider the description relative to one QRF, denoted as $F$. It is also possible to choose $A$ as the reference frame. In that case, $F$ and $A$ are the same quantum system in a measurement~\cite{Brukner, Yang2020}. Fig. 1 shows a schematic illustration of the entities in the relational formulation.

\begin{figure}
\begin{center}
\includegraphics[scale=2.2]{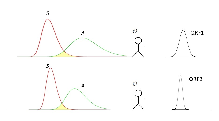}
\caption{Schematic illustration of the entities for the terminologies. The overlapping of the measured system $S$ and apparatus $A$ represents there is interaction in a measurement. The relational properties between $S$ and $A$ must be described relative a QRF. There can be multiple QRFs. Selecting a different QRF, $\cal{O}$ can have a different description of the relational properties in a quantum measurement event.}
\label{fig:1}       
\end{center}
\end{figure}

\subsection{Quantum Measurement}
\label{subsec:quantmeas}
Given the hypothesis that a quantum system should be described relative to another system, the first question to ask is which another system the description is relative to? A quantum system, at any given time, is either being measured by an apparatus, or interacting with its environment, or is in an isolated environment. It is intuitive to select a reference system that has been previously interacting with the quantum system. A brief review of the traditional quantum measurement theory is helpful since it brings important insights on the meaning of a quantum state. 

In the traditional quantum measurement theory proposed by von Neumann\cite{Neumann}, both the quantum system and the measuring apparatus follow the same quantum mechanics laws. Von Neumann further distinguished two separated measurement stages, Process 1 and Process 2. Mathematically, an ideal measurement process is expressed as
\begin{equation}
    \label{measurement}
    \begin{split}
    |\Psi\rangle_{SA} & = |\psi_S\rangle|a_0\rangle = \sum_i c_i|s_i\rangle|a_0\rangle \\
    & \longrightarrow \sum_i c_i|s_i\rangle|a_i\rangle \longrightarrow |s_n\rangle|a_n\rangle
    \end{split}
\end{equation}
Initially, both $S$ and $A$ are in a product state described by $|\Psi\rangle_{SA}$. In Process 2, referring to the first arrow in Eq.(\ref{measurement}), the quantum system $S$ and the apparatus $A$ interact. However, as a combined system they are isolated from interaction with any other system. Therefore, the dynamics of the total system is determined by the Schr\"{o}dinger Equation. Process 2 establishes a one to one correlation between the eigenstate of observable of $S$ and the pointer state of $A$. After Process 2, there are many possible outcomes to choose from. In the next step which is called Process 1, referring to the second arrow in Eq.(\ref{measurement}), one of these possible outcomes (labeled with eigenvalue $n$) emerges out from the rest\footnote{Traditional quantum mechanics does not provide a theoretical description of Process 1. In the Copenhagen Interpretation, this is considered as the ``collapse" of the wave function into an eigenstate of the measured observable. The nature of this wave function collapse has been debated over many decades.}. An observer knows the outcome of the measurement via the pointer variable of the apparatus. Both systems encode information each other, allowing an observer to infer measurement results of $S$ by reading pointer variable of $A$. This observation is also applicable in the case that a quantum system is \textit{prepared} in a particular state. The term preparation refers to the situation that $S$ is measured by an apparatus, or is prepared with a particular lab setup (for instance, a spin half particle passes through a Stern-Gerlach Apparatus), such that its state is explicitly known to an observer. The measuring system, and the environment that $S$ interacts with, are collectively termed as the apparatus $A$. Because of the correlation established between $S$ and $A$ during the state preparation process, it is natural to describe the state of $S$ in reference to $A$. 

After the state preparation, suppose the interaction Hamiltonian between $S$ and $A$ vanishes, $S$ starts its unitary time evolution. During time evolution, $S$ can still be described in reference to the original apparatus $A$. The dynamics is deterministically governed by the Schr\"{o}dinger Equation, but there is no change of correlation between them because there is no interaction. When the next measurement occurs, or when the unitary time evolution stops because $S$ starts to interact with another apparatus $A'$, the relational properties are updated. As a result, the quantum state of $S$ is updated in reference to $A'$. After the interaction finishes, $S$ enters unitary time evolution again. This process can be repeated continuously. 

The key insight of quantum measurement is that it is a question-and-answer bidirectional process. The measuring system interacts (or, disturbs) the measured system. The interaction in turn alters the state of the measuring system. As a result, a correlation is established, allowing the measurement result for $S$ to be inferred from the pointer variable of $A$. 

\subsection{Quantum State}
The notion of information in Ref.~\cite{Rovelli96}, is closely related to concept of correlation. Information exchange between the observed system and the observing apparatus implies change of correlation between the two systems. Correlation is relational and observer-dependent. There are many ways to mathematically define correlation, one of them is introduced in Section \ref{subsec:entanglement}. However, in this paper, we use the notion of information in a more general sense. It can be understood as data that represents values attributed to parameters or properties, or knowledge that describes understanding of physical systems or abstract concepts. Correlation is one type of information.

A \textit{Quantum State} of $S$ describes the complete information an observer $\cal{O}$ can know about $S$. From the examination on the measurement process and the interaction history of a quantum system, we consider a quantum state encodes the information relative to the measuring system or the environment that the system previously interacted with. In this sense, the quantum state of $S$ is described relative to $A$. 
The idea that a quantum state encodes information from previous interactions is also proposed in Ref~\cite{Rovelli07}. The information encoded in the quantum state is the complete knowledge an observer can say about $S$, as it determines the possible outcomes of the next measurement. When the next measurement with another apparatus $A'$ is completed, the description of quantum state is updated to be relative to $A'$.   

In traditional quantum mechanics, the quantum state is described through an observer-independent variable, the wave function $|\psi\rangle$. Its meaning is assigned through the Born's rule~\cite{Dirac58}, which states that the probability to find $S$ in an eigenvector $|s_i\rangle$ is given by $p_i= |\langle s_i|\psi\rangle|^2$, and $\sum p_i=1$. However, in this paper we consider observer-dependent relational properties more basic. By developing a framework to calculate the quantum probability, the meaning of $|\psi\rangle$ is naturally emerged as a secondary concept, as shown in Section \ref{subsec:WF}. 

With the clarifications of the key terminologies, we can proceed to introduce the postulates and start the reformulation of quantum mechanics.

\section{Results}
\label{sec:results}
\subsection{Probability in a Quantum Measurement}
\label{subsec:postulates}
Suppose there is no quantum mechanics formulation yet and the goal is to construct a quantum theory that describes a quantum system $S$ in the context of measurement by an apparatus $A$. We start the reconstruction process by asking a basic question - what are the possible outcomes if one performs a measurement on $S$ using apparatus $A$? More specifically, if one measures a variable $q$ of $S$ by referring a pointer variable $q'$ of $A$, what are the expected outcomes?

To begin with, we assume a fixed QRF, $F$, is chosen to describe the quantum measurement event. From experimental observations, the measurement yields multiple possible outcomes randomly. Each potential outcome is obtained with a certain probability. We call each measurement with a distinct outcome a quantum event. Denote these alternatives events with a set of kets $\{|s_i\rangle\}$ for $S$, where ($i=0,\ldots ,N-1)$, and a set of kets $\{|a_j\rangle\}$ for $A$, where ($j=0,\ldots ,M-1)$. A potential measurement outcome is represented by a pair of kets $(|s_i\rangle, |a_j\rangle)$. The ket $|s_i\rangle$ is introduced not to represent a quantum state of $S$, instead as an abstract notation for a quantum event. They reflect the experimental observations that there can be many distinct measurement outcomes when a variable of $S$, $q$, is measured. $|s_i\rangle$ is associated with one of the outcomes with a certain probability, with $q_i$ as the measured value for variable $q$. Similarly, a ket $|a_j\rangle$ represents a measurement outcome when the pointer variable $q'$ equals $q'_j$. Here finite number of measurement outcomes is assumed. It is always possible to extend the notation to infinite number of events.

With such a representation, the next step is to develop a mathematically framework to calculate the probabilities of possible events. This prediction is carried out before a measurement is actually performed. For instance, what is the probability of a joint event $|s_i\rangle$ and $|a_j\rangle$, denoted as $p_{ij}$? It is subtle to assign a probability of an outcome from a quantum measurement process. As mentioned earlier, a measurement is an inferring process that depends on the physical interaction between $S$ and $A$. The interaction process consists $A$ probing (or, disturbing) $S$, and $S$ in the same time altering $A$. In other words, it is a bidirectional process. We denote this as $A\rightleftharpoons S$. Accordingly, $p_{ij}$ is called an \textit{interactional probability}. This process is true for measurement in either classical or quantum mechanics. The difference is that in classical mechanics, the measurement can be setup such that there is only one measurement outcome deterministically. This means there is a one-to-one correlation between the macroscopic state of measured object and the pointer variable in the measuring device. The probability of this correlation always equals to one. On the other hand, in quantum mechanics, measurement of a variable $q$ of the quantum system $S$ yields multiple possible results. To calculate the probability of the joint event $|s_i\rangle$, and $|a_j\rangle$, the process $A\rightleftharpoons S $ at the macroscopic level should be replaced by $|a_j\rangle \rightleftharpoons |s_i\rangle$ at the quantum level. This is a two-way process, or, a questioning and answering pair in terms of quantum logic approach~\cite{Rovelli96}.  Although the bidirectional interaction process is well known, the following realization is not fully appreciated in the research literature.
\begin{displayquote}
\textit{Because a quantum measurement is a bidirectional process, the calculation of the probability of a measurement outcome must faithfully model such bidirectional process.}
\end{displayquote}
The bidirectional process does not necessarily imply two sequential steps. Instead, the probing and responding processes are understood as two aspects of a complete process in a measurement event. We can use a classical probability problem to analogize this. Suppose tossing a special coin gets a face up with probability of $p$. Let us consider a measurement process that requires tossing two such coins in the same time, and the measurement is successful if one coin facing up and one coin facing down. We ask what is the probability of a successful measurement event. The answer is to multiple two probability quantities together, $p(1-p)$. In the similar manner, given the bidirectional process in a quantum measurement event, the observable measurement probability should be calculated as a product of two quantities of weights. One weight quantity is associated with the probing process from $A\to S$, denoted as $Q^{A\rightarrow S} (|a_j\rangle \cap |s_i\rangle)$, and the other is associated with the responding process from $S\to A$, denoted as $R^{S\rightarrow A} (|s_i\rangle \cap |a_j\rangle)$, so that
\begin{equation}
\label{prob1}
    p_{ij} \propto Q^{A\rightarrow S} (|a_j\rangle \cap |s_i\rangle)R^{S\rightarrow A} (|s_i\rangle \cap |a_j\rangle)
\end{equation}
Here we assume process of each direction is independent from each other. The requirements for the interactional probability $p_{ij}$ can be summarized as following:
\begin{enumerate}
\item $p_{ij}$ should be a product of two numbers that are associated with a bidirectional process.
\item $p_{ij}$ should be symmetric with respect to either $S$ or $A$. What this means is that the probability is the same for both processes $|a_j\rangle \rightarrow |s_i\rangle \rightarrow |a_j\rangle$ that is viewed from $A$ and $|s_i\rangle \rightarrow |a_j\rangle \rightarrow |s_i\rangle$ that is viewed from $S$.
\item $p_{ij}$ should be a non-negative real number.
\end{enumerate}
To satisfy requirement 2, we rewrite these two quantities as matrix elements, i.e., $Q^{A\rightarrow S} (|a_j\rangle \cap |s_i\rangle) = Q^{AS}_{ji}$, and $R^{S\rightarrow A} (|s_i\rangle \cap |a_j\rangle) = R^{SA}_{ij}$. Eq.(\ref{prob1}) becomes 
\begin{equation}
\label{probprod}
    p_{ij} \propto Q_{ji}^{AS}R_{ij}^{SA}. 
\end{equation}
The probability for the process $|s_i\rangle \rightarrow |a_j\rangle \rightarrow |s_i\rangle$ is $p_{ij} \propto R_{ij}^{SA}Q_{ji}^{AS}$, the same as Eq.(\ref{probprod}). Thus, requirement 2 is satisfied.

Now let's consider the third requirement for $p_{ij}$ that it should be a non-negative real number. We should assume the weakest possible restrictions to the variables $Q_{ji}^{AS}$ and $R_{ij}^{SA}$. The three requirements for $p_{ij}$ are not necessarily applicable to $Q_{ji}^{AS}$ and $R_{ij}^{SA}$. First, a unidirectional process $|a_j\rangle \rightarrow |s_i\rangle$ does not constitute a complete physical measurement process. We should not consider these variables themselves as probability quantities in the classical sense. This is, $Q_{ji}^{AS}$ and $R_{ij}^{SA}$ are not necessarily non-negative real number. They can be complex numbers. In other words, a probabilistic quantity is a non-negative real number only when it is associated with an actual physical measurement. Such a requirement does not need to be true for probabilistic quantity associated with an incomplete, one-way process. A similar argument can be found in Ref.~\cite{Feynman48}. We summarize this crucial but subtle point as following:
\begin{displayquote}
\textit{Measurement probability of an observable event must be a non-negative real number. However, the requirement of being a non-negative real number is not applicable to non-measurable quantities for sub-processes that constitute a quantum measurement.}
\end{displayquote}
There is also a temptation to express the relational variable as $Q^{A\rightarrow S} (|a_j\rangle \mid |s_i\rangle)$, making it looks like a conditional probability quantity. However, we choose the expression $Q^{A\rightarrow S} (|a_j\rangle \cap |s_i\rangle)$ because it better represents a relational quantity for a joint event. 

Second, there is no reason to assume $R^{SA}_{ij} = Q^{AS}_{ji}$. The direction from $S$ to $A$ is significant here and explicitly called out in the superscript\footnote{In this notation, index $i$ is reserved for $S$ while index $j$ is reserved for $A$.}. To see this, let us analyze the factors that the weight $Q^{A\rightarrow S} (|a_j\rangle \cap |s_i\rangle)$ depends on. Intuitively, this quantity depends on three factors:
\begin{enumerate}
\item Likelihood of finding system S in state $|s_i\rangle$ without interaction;
\item Likelihood of finding system A in state $|a_j\rangle$ without interaction;
\item A factor that alters the above two likelihoods due to the passing of physical elements such as energy and momentum from $A\to S$ in the probing process.
\end{enumerate}
Similarly, the other weight quantity $R^{S\rightarrow A} (|s_i\rangle \cap |a_j\rangle)$ depends on the similar first two factors, and the third factor that is due to the passing of physical elements from $S\to A$ in the responding process.

The three dependent factors for $Q_{ji}^{AS}$ and $R_{ij}^{SA}$ are related to each other respectively. The likelihood of finding system $S$ in state $|s_i\rangle$ and system $A$ in state $|a_j\rangle$ without interaction are the same. The third factor is triggered by passing physical elements during interaction. There are conservation laws such as energy conservation and momentum conservation during interaction. Conceivably, the third factors for $Q_{ji}^{AS}$ and $R_{ij}^{SA}$ must be equal in absolute value, but may be different in phase. With all these considerations, it is reasonable to assume $|Q_{ji}^{AS}| = |R_{ij}^{SA}|$. We choose
\begin{equation}
\label{conjugate}
    Q_{ji}^{AS} = (R_{ij}^{SA})^*
\end{equation}
so that $p_{ij}=Q_{ji}^{AS}R_{ij}^{SA}$ is a non-negative real number. Written in a different format, $Q_{ji}^{AS} = (R^{SA})^\dag_{ji}$. This means $Q^{AS} = (R^{SA})^\dag$. Eq.(\ref{probprod}) then becomes 
\begin{equation}
    p_{ij} = |R^{SA}_{ij}|^2/\Omega
\end{equation}
where $\Omega$ is a real number normalization factor. Eq.(\ref{conjugate}) can be intuitively understood as this: viewed from $A$ or viewed from $S$, the probabilistic quantities have the same magnitude, but different in phase. Physically it ensures there is no preferred choice of $S$ and $A$ in defining the relational variables\footnote{When the correlation between $S$ and $A$ are established, both systems are effectively measuring each other (see similar remark in Ref~\cite{Zurek03}). Change occurs either in $S$ or in $A$ will be reflected by the relational matrix element. But there should not have a preference of considering $S$ or $A$ as a measuring system.}. Eq.(\ref{conjugate}) is a weaker version of requirement for $R^{AS}_{ij}$ compared to the second requirement for $p_{ij}$.  $Q_{ji}^{AS}$ and $R_{ij}^{SA}$ are called \textit{relational probability amplitudes}. In Section \ref{sec:PI}, we will give an explicit calculation of $R^{SA}_{ij}$, using the Path Integral formulation. Given the relation in Eq.(\ref{conjugate}), we will not distinguish the notation $R$ versus $Q$, and only use $R$. 

The relational matrix $R^{SA}$ gives the complete description of $S$. It provides a framework to derive the probability of future measurement outcome. We summarize the ideas presented in this section with the following two postulates. 

\begin{displayquote}
\textbf{Postulate 1} \textit{A quantum system $S$ is completely described by a matrix $R^{SA}$ relative to an apparatus $A$, where the matrix element $R^{SA}_{ij}$ is the relational probability amplitude for the joint events $|s_i\rangle$ and $|a_j\rangle$.}
\end{displayquote}
\begin{displayquote}
\textbf{Postulate 2} \textit{Probability of a measurement outcome is calculated by modeling the potential interaction process, i.e., by multiplying two relational probability amplitudes representing the bidirectional process.}
\end{displayquote}
There are two important notes. First, $R^{SA}_{ij}$ is probabilistic quantity, not a quantity associated with certain physical variable. $R^{SA}_{ij}$ should not be considered as certain coupling strength between $S$ and $A$. In Section \ref{sec:PI}, in the context of path integral,  $R^{SA}_{ij}$ is defined as the sum of quantity $e^{iS_p/\hbar}$, where $S_p$ is the action of the composite system $S+A$ along a path. Physical interaction between $S$ and $A$ may cause change of $S_p$, which is the phase of the probability amplitude. But $e^{iS_p/\hbar}$ itself is a probabilistic quantity. Second, although $R^{SA}_{ij}$ is a probability amplitude, not a probability real number, we assume it follows certain rules in the classical probability theory, such as multiplication rule, and sum of alternatives in the intermediate steps.

\subsection{Wave Function and Born's Rule}
\label{subsec:WF}
So far, we have not yet introduced the notion of quantum state for $S$. We only describe $S$ and $A$ with sets of events and the relational matrix $R^{SA}$. The next step is to derive the properties of $S$ based on $R^{SA}$. This can be achieved by examining how the probability of measuring $S$ with a particular outcome of variable $q$ is calculated. 

We will take a move on mathematical notation before proceeding further. It is more convenient to introduce a Hilbert space for the quantum system $S$. The set of kets $\{|s_i\rangle\}$, previously considered as representing distinct measurement events for $S$, can be considered as \textit{eigenbasis} of Hilbert space ${\cal H}_S$ with dimension $N$, and $|s_i\rangle$ is an eigenvector. Since each measurement outcome is distinguishable, $\langle s_i|s_j\rangle = \delta_{ij}$. Similarly, the set of kets $\{|a_j\rangle\}$ is eigenbasis of Hilbert space ${\cal H}_A$ with dimension $N$ for the apparatus system $A$. The bidirectional process $|a_j\rangle \rightleftharpoons |s_i\rangle$ is called a \textit{potential measurement configuration}. A potential measurement configuration comprises possible eigen-vectors of $S$ and $A$ that involve in the measurement event, and the relational weight quantities. It can be represented by $\Gamma_{ij}: \{|s_i\rangle, |a_j\rangle, R^{SA}_{ij}, Q^{AS}_{ji}\}$.

In the previous section, we argue that the probability of the joint events $|s_i\rangle$ and $|a_j\rangle$ is given by $p_{ij} = Q_{ji}^{AS}R_{ij}^{SA}=|R_{ij}^{SA}|^2$, because the corresponding measurement configuration is $|a_j\rangle \rightarrow |s_i\rangle \rightarrow |a_j\rangle$. Here we clearly specify that the probability is for the joint event $|s_i\rangle$ and  $|a_j\rangle$. But there is a limitation for such specification if we wish to calculate the probability of measuring $S$ with a particular outcome of variable $q$. In such case, the measurement configuration used earlier $|a_j\rangle \rightarrow |s_i\rangle \rightarrow |a_j\rangle$ \textit{over-describe} the configuration because no measurement is actually performed. We do not know that which event will occur to the quantum system $A$ since it is completely probabilistic. The only way an observer can determine which event occurs is to perform actual measurement, or to infer from another system. Therefore, it is legitimate to generalize the potential measurement configuration as $|a_j\rangle \rightarrow |s_i\rangle \rightarrow |a_k\rangle$. In other words, the measurement configuration in the joint Hilbert space starts from $|a_j\rangle$, but can end at $|a_j\rangle$, or any other event, $|a_k\rangle$. Correspondingly, we generalize Eq.(\ref{probprod}) by introducing a quantity for such configuration
\begin{equation}
\label{micAction}
W_{jik}^{ASA} = Q^{AS}_{ji}R^{SA}_{ik} = (R^{SA}_{ij})^*R^{SA}_{ik}.
\end{equation}
The second step utilizes Eq.(\ref{conjugate}). We interpret this quantity as a weight associated with the potential measurement configuration $|a_j\rangle \rightarrow |s_i\rangle \rightarrow |a_k\rangle$. The probability for a measurement outcome can be calculated by identifying the appropriate alternatives and summing up their weights. The classical macroscopic configuration $A\rightarrow S\rightarrow A$ can be considered as a special case when the dimension of the Hilbert space is one for either $S$ or $A$. Indeed, the most general form of measurement configuration in a bipartite system can be $|a_j\rangle \rightarrow |s_m\rangle \rightarrow |s_n\rangle \rightarrow |a_k\rangle$, and its weight is given by 
\begin{equation}
\label{micAction2}
W_{jmnk}^{ASSA}=Q^{AS}_{jm}R^{SA}_{nk}. 
\end{equation}
The indeterminacy on which event will occur to a quantum system influences the way possible measurement configurations can be arranged. Consequently, it influences how the applicable configurations are counted and then how the probability is calculated\footnote{The situation when inference information is available is discussed in Section \ref{subsec:entanglement}. In probability theory, it is crucial not to under-count or over-count applicable alternatives when calculating probability. When a quantum system is in a superposition state, although each eigenvector is labeled with a different ket, each ket should be considered indistinguishable for counting purpose because there is no information to determine exactly which ket the system is in. It is an under-count if only considering $|a_j\rangle \rightarrow |s_i\rangle \rightarrow |a_j\rangle$. There is similar example in statistical physics. When counting the number of microscopic states of an ensemble consisting of identical particles, one strategy is to first over-count by assuming the particles are distinguishable, then divide the counting result by a factor to offset the over-counting.}. Suppose we do not perform actual measurement and inference is not available, the probability of finding $S$ in a future measurement outcome can be calculated by summing $W_{jmnk}^{ASSA}$ from all applicable alternatives of measurement configurations. Such generalized framework of calculating probability is stated by extending Postulate 2.
\begin{displayquote}
\textbf{Postulate 2e} \textit{Probability of a measurement outcome is calculated by modeling the potential interaction process. The probability is proportional to the sum of weights from all applicable measurement configurations, where the weight is defined as the product of two relational probability amplitudes corresponding to the configuration.}
\end{displayquote}
With this framework, the remaining task to calculate the probability is to correctly count the \textit{applicable} alternatives of measurement configuration. This task depends on the expected measurement outcome. Some typical cases are analyzed next.

\textit{Case 1.} Suppose the expected outcome of an ideal measurement is event $|s_i\rangle$, i.e., measuring variable $q$ gives eigenvalue $q_i$. The probability of event $|s_i\rangle$ occurs, $p_i$, is proportional to the summation of $W_{jmnk}^{ASSA}$ from all the possible configurations related to $|s_i\rangle$. Mathematically, we select all $W_{jmnk}^{ASSA}$ with $m=n=i$, sum over index $j$ and $k$, and obtain the probability $p_i$.
\begin{equation}
\label{probability}
p_i \propto \sum_{j,k=0}^M (R^{SA}_{ij})^*R^{SA}_{ik}.
\end{equation}
To see why this quantity can be considered a probability number, we note that Eq.(\ref{probability}) is symmetric with respect to the swap of index ${j \leftrightarrow k}$. It can be rewritten as
\begin{equation}
\label{probability2}
p_i \propto \sum_{j} (R^{SA}_{ij})^* \sum_k R^{SA}_{ik}= |\sum_{j} R^{SA}_{ij}|^2.
\end{equation}
It is a positive real number. Normalization condition is given by
\begin{equation}
\label{normalization}
\begin{split}
 \sum_i |\sum_{j} R_{ij}|^2 & =\sum_{jk}\sum_{i} R_{ij}R_{ik}^*\\
 & =\sum_{jk}(R^\dag R)_{jk} = 1.
\end{split}
\end{equation}
A notation move is made in the above equation by omitting the superscript in $R^{SA}$, with the convention that $R$ refers to the relational matrix from $S$ to $A$. The definition of the wave function naturally emerges out from Eq. (\ref{probability2}). Define a variable $\varphi_i = \sum_{j} R_{ij}$, then $p_i=|\varphi_i|^2$. The quantum state can be described either by the relational matrix $R$, or by a set of variables $\{\varphi_i\}$. We call $\varphi_i$ the wave function for eigenvector $|s_i\rangle$. The quantum state of $S$ is a vector representation of the variable set $\{\varphi_i\}$, i.e., the vector state of $S$ relative to $A$, is $|\psi\rangle_S^A = (\varphi_0, \varphi_1,\ldots, \varphi_N)^T$ where superscript $T$ is the transposition symbol. In summary,
\begin{equation}
\label{WFT}
|\psi\rangle_S^A = (\varphi_0, \varphi_1,\ldots, \varphi_N)^T  \quad \textrm{where } \varphi_i = \sum_{j} R_{ij}.
\end{equation}
Note that we have not yet written $|\psi\rangle_S^A$ as linear combination of $\varphi_i$. 

\textit{Case 2.} Suppose the expected ideal measurement outcome is that $S$ in a superposition of eigenvectors $|s_0\rangle$ and $|s_1\rangle$. This means one cannot determine event $|s_0\rangle$ or $|s_1\rangle$ occurs. The compute the probability, the applicable weights should include not only $\sum_{jk}R_{0j}^*R_{0k}=|\sum_jR_{0j}|^2=|\varphi_0|^2$ and $\sum_{jk}R_{1j}^*R_{1k}=|\sum_jR_{1j}|^2=|\varphi_1|^2$, but also the terms that index 0 and 1 are inter-exchanged due to the indeterminacy, i.e., $\sum_{jk}R_{0j}^*R_{1k}$ and $\sum_{jk}R_{1j}^*R_{0k}$. Adding these terms together, the probability is
\begin{equation}
\label{probability4}
p_{\{0,1\}}= |\sum_jR_{0j} + \sum_jR_{1j}|^2 = |\varphi_0 + \varphi_1|^2
\end{equation}
Eq.(\ref{probability4}) captures the characteristics of superposition. The wave function for a superposition of eigenvectors $|s_0\rangle$ and $|s_1\rangle$ is the linear combination of $\varphi_0$ and $\varphi_1$. Based on this observation, it is mathematically convenient to write the state vector of $S$ as linear combination of $\varphi_i |s_i\rangle$
\begin{equation}
\label{WF}
|\psi\rangle_S^A = \sum_i\varphi_i |s_i\rangle  \quad \textrm{where } \varphi_i = \sum_{j} R_{ij}.
\end{equation}
The justification for the above definition is that the probability can be calculated from it by defining a projection operator $\hat{P}_i=|s_i\rangle\langle s_i|$. Noted that $\{|s_i\rangle\}$ are orthogonal eigenbasis, the probability is rewritten as:
\begin{equation}
\label{probability3}
p_i=\langle\psi|\hat{P}_i|\psi\rangle = |\varphi_i|^2
\end{equation}
Similarly, introducing a projection operator $\hat{P}_{\{0,1\}}=(|s_0\rangle+|s_1\rangle)(\langle s_0|+\langle s_1|)$, we can rewrite the probability as
\begin{equation}
\label{probability5}
p_{\{0,1\}}=\langle\psi|\hat{P}_{\{0,1\}}|\psi\rangle = |\varphi_0 + \varphi_1|^2.
\end{equation}
Eq.(\ref{WF}) and (\ref{probability3}) give the equivalent results as what Born's Rule states, but with more physical insights on how the quantum measurement probability is calculated based on detailed analysis on the interaction process during measurement. 

\textit{Case 3.} Given a relational matrix $R$ and that the correspondent state vector of $S$ is $|\psi\rangle$, suppose the expected measurement outcome is described by another relational matrix $Q$ and the correspondent state vector of $S$ is $|\chi\rangle$, the probability is 
\begin{equation}
    \label{generalizedProb}
    p(Q|R) =  \| \sum_{i,j}(Q^\dag R)_{i,j} \|.
\end{equation}  
The poof is given in Section \ref{subsec:generalizedProb}. Using the state vector notation of $S$, the probability can be equivalently expressed as $p(\chi | \psi)  = \langle\psi|\hat{P}_{\chi}|\psi\rangle = \| \langle\chi | \psi\rangle \|$, where $\hat{P}_{\chi} = |\chi\rangle\langle\chi |$. This is a generalization of Eq.(\ref{probability5}).  

Although the introduction of wave function $\varphi_i$ brings much mathematical convenience, the relational matrix $R$ is a more fundamental quantity. $\varphi_i$ is introduced as a byproduct of the derivation instead of as a fundamental variable.

Eqs.(\ref{probability}) and (\ref{WF}) are introduced on the condition that there is no correlation between quantum system $S$ and $A$. If there is correlation between them, the summation in Eq.(\ref{probability}) over-counts the applicable alternatives of measurement configurations and should be modified accordingly. But first, from the relational matrix $R$, how to determine whether there is a correlation between $S$ and $A$?

\subsection{Entanglement}
\label{subsec:entanglement}

Correlation between two quantum system means one can infer the information on one system from information on the other system. The relational variable $R_{ij}$ itself does not quantify an inference relation between $S$ and $A$. Quantity $|R_{ij}|^2$ is the measurement probability for the joint events $|s_i\rangle$ for $S$ \textit{and} $|a_j\rangle$ for $A$. However, given $R_{ij}$, one cannot infer that event $|s_i\rangle$ occurs to $S$ from knowing event $|a_j\rangle$ occurs to $A$.  We need to define a different parameter that can quantify the quantum correlation between $S$ and $A$. 

The capability of inferring information of a quantum state of one system from information of the other system is defined as entanglement. Since $S$ and $A$ both are quantum systems, they form a bipartite quantum system. Entanglement between two composite system is quantified by an entanglement measure $E$. There are many forms of entanglement measure~\cite{Nelson00, Horodecki}, the simplest one is the von Neumann entropy. Given the relational matrix $R$, the von Neumann entropy is defined as following. For reason that will become obvious in Section \ref{subsec:2ndO}, we first define a product matrix $\rho = RR^\dag$, and denote the eigenvalues of $\rho$ as $\{\lambda_i\}$, then the von Neumann entropy for the relational matrix $R$ is
\begin{equation}
    \label{vonNeumann}
    H(R)  = -\sum_i\lambda_i ln\lambda_i.
\end{equation}
A change in $H(R)$ implies there is change of entanglement between $S$ and $A$. Unless explicitly pointed out, we only consider the situation that $S$ is described by a single relational matrix $R$. In this case, the entanglement measure $E=H(R)$.

The definition of $H(R)$ enables us to distinguish different quantum dynamics. Given a quantum system $S$ and its referencing apparatus $A$, there are two types of the dynamics between them. In the first type of dynamics, there is no physical interaction and no change in the entanglement measure between $S$ and $A$. $S$ is not necessarily isolated in the sense that it can still be entangled with $A$, but the entanglement measure remains unchanged. This type of dynamics is defined as \textit{time evolution}. In the second type of dynamics, there is a physical interaction and correlation information exchange between $S$ and $A$, i.e., the von Neumann entropy $H(R)$ changes. This type of dynamics is defined as \textit{quantum operation}. \textit{Quantum measurement} is a special type of quantum operation with a particular outcome. Whether the entanglement measure changes distinguishes a dynamic as either a time evolution or a quantum operation. This is summarized in the following postulate.
\begin{displayquote}
\textbf{Postulate 3} \textit{In a time evolution process, the entanglement measure of relational matrix is unchanged, while in a quantum operation process, there is change in the entanglement measure of relational matrix.}
\end{displayquote}
The following theorem allows us to detect whether relational matrix $R$ is entangled. The theorem will be used extensively later.
\begin{theorem} 
\label{productStateTheo}
$H(R)=0$ if and only if the matrix element $R_{ij}$ can be decomposed as $R_{ij}=c_id_j$, where $c_i$ and $d_j$ are complex numbers.
\end{theorem}
The proof is left to the Section \ref{subsect:theorem1}. The wave function in this case is simplified to $\varphi_i=\sum_j c_id_j=c_i\sum_jd_j=c_id$ where $d$ is a constant. If we choose $\sum_i|c_i|^2=1$, then $d=e^{i\phi}$. For simplicity, let $d=1$ so that $\varphi_i=c_i$. 

When there is entanglement between $S$ and $A$, $A$ and $S$ can infer information from each other. The way probability is calculated in Eqs.(\ref{probability}) and (\ref{probability4}) must be modified because the summation in Eq.(\ref{probability}) over counts the alternatives that are due to indeterminacy. Some of the potential measurement configurations should be excluded in order to calculate the probability correctly. 

To see this more clearly, we decompose the relational matrix $R$ to $R=UDV$ by virtue of the singular value decomposition, where $U$ and $V$ are two unitary matrices, and $D$ is a diagonal matrix. Applying the two unitary matrices is equivalent to changing eigenbasis of $S$ and $A$ to $|\tilde{s}_i\rangle$ and $|\tilde{a}_i\rangle$ such that $R$ is diagonal. $D$ is an irreducible diagonal matrix. $H(R)>0$ implies that $D$ has more than one diagonal matrix elements. Each element corresponds to a one to one correlation between $|\tilde{s}_i\rangle$ and $|\tilde{a}_i\rangle$. One can infer $S$ is in $|\tilde{s}_i\rangle$ from knowing $A$ is in $|\tilde{a}_i\rangle$, and vice versa. Effectively, neither $S$ nor $A$ is in a superposition state anymore. The contributions in calculating probability due to indeterminacy of eigenvectors must be excluded. This results in a different rule to count the applicable alternatives.

\textit{Case 1e.} When there is an entanglement between $S$ and $A$, to calculate the probability of finding $S$ in eigenvector $|s_i\rangle$, one should only select measurement configuration $|a_j\rangle \rightarrow |s_i\rangle \rightarrow |a_j\rangle$. The corresponding weight is $R_{ij}^*R_{ij}=|R_{ij}|^2$. Summing all possible index $j$ give the probability
\begin{equation}
    \label{probability6}
    p_i=\sum_j|R_{ij}|^2
\end{equation}

\textit{Case 2e.} Suppose we want to calculate the probability of finding $S$ in eigenvectors $|s_0\rangle$ or $|s_1\rangle$ when there is an entanglement between $S$ and $A$. Since there is inference information on whether $S$ is in eigenvector $|s_0\rangle$ or $|s_1\rangle$, to calculate the probability, we can only count the weights $\sum_jR^*_{0j}R_{0j}$ and $\sum_jR^*_{1j}R_{1j}$, and not to include the interference terms such as $\sum_jR^*_{0j}R_{1j}$.
\begin{equation}
    \label{probability7}
    p_{\{0,1\}}=\sum_j|R_{0j}|^2 + \sum_j|R_{1j}|^2 = p_0+p_1.
\end{equation}

It is worth to mention that the rule to calculate the probability when there is an entanglement between $S$ and $A$ echoes the idea from Feynman's path integral formulation. In path integral formulation~\cite{Feynman05}, the probability amplitude is calculated by summing possible alternatives. When the alternatives can be indistinguishable, they are called ``interfering alternatives". When the alternatives are distinguishable such as in the case of entanglement between $S$ and $A$, they are called ``exclusive alternatives". Supposed each alternative is assigned a weight, the rules to calculate the probability can be summarized as
\begin{displayquote}
\textbf{Probability for Alternatives} \textit{To calculate the probability for interfering alternatives, one first takes the summation of weight for each alternative, then takes the modulus square of the summation. To calculate the probability for exclusive alternatives, one first takes the modulus square of weight for each alternative, then takes the summation.}
\end{displayquote}
In the other words, the order of taking modulus square and taking summation is swapped for both cases, as clearly shown in (\ref{probability2}) and (\ref{probability6}). This rule is introduced as a postulate in path integral~\cite{Feynman05} and the justification for the step of taking modulus square is not provided. Here we complete the justification by explaining that the modulus square is due to the bidirectional measurement process, as shown in the derivation of (\ref{probability2}).

As a consequence of entanglement, we cannot define a wave function as in Eq.(\ref{WF}) to describe the state of $S$. To describe $S$ without explicitly referencing $A$ when $S$ and $A$ are entangled, alternative formulation is needed. This is the reduced density matrix approach.

\subsection{Reduced Density Matrix}
\label{subsec:2ndO}
To describe $S$ without explicitly referencing $A$ when $S$ and $A$ are entangled, we first describe the composite system $S+A$ as an isolated system such that Eq.(\ref{WF}) is applicable. We need to describe $S+A$ relative to another measurement apparatus $A'$ that is unentangled with $S+A$. Suppose an observer $\cal{O}_E$ is local to apparatus $A'$, and has the same information of the relational matrix $R$ and using the same reference frame $F$. $\cal{O}_E$ wishes to describe the composite system $S+A$ using Postulate 2e. In order to describe a quantum state of a composite system, another postulate is needed, which is commonly found in standard textbooks, for example,
\begin{displayquote}
\textbf{Postulate 4} \textit{Let $S_{12}$ be the composite system of quantum system $S_1$ and $S_2$ with Hilbert spaces ${\cal H}_1$ and ${\cal H}_2$. Then the associated Hilbert space of $S_{12}$ is a tensor product Hilbert space ${\cal H}_1\otimes {\cal H}_2$. A physical variable of $S_1$ represented by Hermitian operator $A_1$ on ${\cal H}_1$ is identified with the physical variable of $S_{12}$ represented by $A_1\otimes I_2$ on ${\cal H}_1\otimes {\cal H}_2$, where $I_2$ is the identity operator on ${\cal H}_2$~\cite{Hayashi15}.}
\end{displayquote}
An eigenvector denotes a distinct quantum event that a measurement of a variable yield a distinct eigenvalue. If there are $N$ orthogonal eigenvectors for $S$, $\{|s_i\rangle\}$, and $M$ orthogonal eigenvectors for $A$, $\{|a_i\rangle\}$, according to Postulate 4, the orthogonal basis set for the composite $S+A$ system should have $N\times M$ eigenvectors, $\{|s_i\rangle\otimes |a_j\rangle\}$. $\cal{O}_E$ would describe $S+A$ with a higher order relational matrix, denoted as $R'$, with matrix element $R'_{mn}$. Index $m$ is defined in Hilbert space ${\cal H}_S\otimes {\cal H}_A$, $(m=0, \ldots ,NM-1)$, while index $n$ is defined in Hilbert space ${\cal H}_{A'}, (n=0, \ldots ,M'-1)$ and $M'=\dim {\cal H}_{A'}$. Since there is no entanglement between $A'$ and $S+A$, $R'$ can be used to define the wave function of the composite system as $\varphi_m^{A'} = \sum_{n} R^{\prime}_{mn}$ according to Eq.(\ref{WF}). However, there is restrictions on $R^{\prime}_{mn}$ because the relational matrix between $S$ and $A$ has been established. The relational matrix $R'$ must satisfy the following condition\footnote{From Postulate 2e, the probability of finding the composite system $S+A$ in an eigenvector $|m\rangle$ is $p_m=|\sum_{n} R^{\prime}_{mn}|^2$. From Postulate 4, $|m\rangle$ can be rewritten to be $|s_i\rangle|a_j\rangle$ by renumbering index $m$ to $i,j$ since $m$ is defined in the Hilbert space ${\cal H}_S\otimes {\cal H}_A$. Therefore $p_m$ is the probability for the combined events $|s_i\rangle$ for $S$ and $|a_j\rangle$ for $A$, i.e., $p_m=p_{ij}$. But $p_{ij}=|R_{ij}|^2$ so that $|\sum_{n} R^{\prime}_{mn}|^2=|R_{ij}|^2$. This gives $\sum_{n} R^{\prime}_{mn}=e^{i\phi}R_{ij}$ where $e^{i\phi}$ is an unimportant phase factor.} 
\begin{equation}
\label{WF2}
    \varphi_m=\sum_{n} R^{\prime}_{mn} =R_{ij}.
\end{equation}
Therefore, relative to $\cal{O}_E$, the state vector of the composite system $S+A$ is
\begin{equation}
\label{WF3}
\begin{split}
|\Psi\rangle & = \sum_m \varphi_m|m\rangle = \sum_{ij} \varphi_{ij} |s_i a_j\rangle \\
& = \sum_{ij}R_{ij}|s_i\rangle|a_j\rangle.
\end{split}
\end{equation}
Next, we ask how to describe $S$ itself. To answer this, we examine how the probability of the system $S$ in an eigenvector $|s_i\rangle$ can be calculated. We know that the probability of $S$ in eigenvector $|s_i\rangle$ \textit{and} $A$ in eigenvector $|a_j\rangle$ is $p_{ij}=|R_{ij}|^2$. If event 1.)$S$ in eigenvector $|s_i\rangle$ \textit{and} $A$ in eigenvector $|a_j\rangle$, and event 2.)$S$ in eigenvector $|s_i\rangle$ \textit{and} $A$ in eigenvector $|a_k\rangle$, are mutually exclusive, the probability of $S$ in eigenvector $|s_i\rangle$ is then just the sum of $p_{ij}$ over index $j$, i.e., $p_i=\sum_ip_{ij}=\sum_j|R_{ij}|^2$. It gives the desired result as Eq.(\ref{probability6}) when $S$ and $A$ are entangled. This is not a surprise since the assumption that event-1 and event-2 are mutually exclusive implies there is no event such that $S$ is in eigenvector $|s_i\rangle$ while $A$ is in eigenvector $|a_j\rangle$ \textit{and} $|a_k\rangle$. In other words, the mutual exclusivity of event-1 and event-2 eliminates the potential measurement configuration $|a_j\rangle \rightarrow |s_i\rangle \rightarrow |a_k\rangle$, thus satisfies the requirement for calculating probability when there is entanglement between $S$ and $A$. The mathematical tool to implement this requirement is the reduced density operator of $S$, defined as
\begin{equation}
    \label{density}
    \begin{split}
    \hat{\rho}_S & =Tr_A|\Psi\rangle\langle\Psi| = \sum_{ii'}(\sum_kR_{ik}R^*_{i'k})|s_i\rangle\langle s_{i'}| \\
    & =\sum_{ii'}(RR^\dag)_{ii'}|s_i\rangle\langle s_{i'}|.
    \end{split}
\end{equation}
The partial trace over $A$, $Tr_A(.) = \sum_k\langle a_k|.|a_k\rangle$, ensures the mutual exclusivity of event-1 and event-2 since only the diagonal elements are selected in the sum. To obtain the desired probability $p_i=\sum_j|R_{ij}|^2$, we define a projection operator $\hat{P}_i=|s_i\rangle\langle s_i|$, so that
\begin{equation}
\label{indirectProb}
p_i = Tr_S(\hat{P}_i\hat{\rho}_S) = \sum_j|R_{ij}|^2.
\end{equation}
Since the information of $A$ is traced out in $\hat{\rho}_S$, we find a mathematical tool to describe the state of $S$ without explicitly referring to $A$. Eq.(\ref{density}) gives a clear meaning of the matrix product $RR^\dag$ as the reduced density matrix of $S$, i.e., $\rho_S=RR^\dag$. Thus, the entanglement measure defined in (\ref{vonNeumann}) is the von Neumann entropy for the reduced density matrix of $S$. 

Similarly, the probability of event $|a_j\rangle$ for $A$ is $p^A_j=\sum_ip_{ij}=\sum_i|R_{ij}|^2$. This can be more elegantly written by introducing a partial projection operator $I^S\otimes \hat{P}_j^A$ where $\hat{P}_j^A=|a_j\rangle\langle a_j|$. It is easy to verify that
\begin{equation}
\label{partialProb}
\begin{split}
p_j^A & =\langle\Psi|I^S\otimes \hat{P}_j^A|\Psi\rangle \\
& =\langle\Psi|a_j\rangle\langle a_j|\Psi\rangle =\sum_i|R_{ij}|^2
\end{split}
\end{equation}

To calculate the probability of finding $S$ in $|s_0\rangle$ or $|s_1\rangle$, we use the projection operator defined as $\hat{P}_{\{01\}}=|s_0\rangle\langle s_0|+|s_1\rangle\langle s_1|$, then 
\begin{equation}
\label{indirectProb2}
\begin{split}
p_{\{0,1\}} &= Tr_S(\hat{P}_{\{01\}}\hat{\rho}_S) \\
& = \sum_j|R_{0j}|^2 + \sum_j|R_{1j}|^2 = p_0+p_1.
\end{split}
\end{equation}
which is the same as Eq. (\ref{probability7}) in \textit{Case 2e}. The trace operation over $S$ in Eqs. (\ref{indirectProb}) and (\ref{indirectProb2}) takes only diagonal matrix elements, effectively eliminates the indeterminacy with respect to eigenvector $|s_i\rangle$ in the information acquisition flows. Together with the partial trace operation in the definition of $\hat{\rho}_S$, they exclude the interference terms in the summation of weights for calculation of probability, thus effectively factor in the inference information between $S$ and $A$, and yield the same results as deduced from Postulate 2e.  

Normalization of $|\Psi\rangle$ requires
\begin{equation}
    \label{normalization2}
    \begin{split}
    Tr(\rho_S) & = Tr(RR^\dag)= \sum_i(\sum_jR_{ij}R^\dag_{ji}) \\
    & = \sum_{ij}|R_{ij}|^2 =1
    \end{split}
\end{equation}
Note Eqs.(\ref{normalization}) and (\ref{normalization2}) cannot be true in the same time. Eq.(\ref{normalization}) is true only when the relational matrix $R$ is unentangled. When $S+A$ is entangled, Eq.(\ref{WF}) cannot be used to describe $S$. This is evident if we rewrite Eq.(\ref{WF}) in the density matrix operator format, 
\begin{equation}
    \label{density1}
    \begin{split}
    \hat{\rho}_S^{\prime} & = |\psi\rangle_{S}\langle\psi|= \sum_{ii'}(\sum_{jj'}R_{ij}R^*_{i'j'})|s_i\rangle\langle s_{i'}|\\
    & = \hat{\rho}_S + \sum_{ii'}(\sum_{j\ne j'}R_{ij}R^*_{i'j'})|s_i\rangle\langle s_{i'}|.
    \end{split}
\end{equation}
Clearly, $\hat{\rho}_S^{\prime} \ne \hat{\rho}_S$ in general. The second term in Eq.(\ref{density1}) comes from the indeterminacy of eigenvector for $A$. This term should be discarded when $H(R)>0$. This confirms that $S$ should be described by Eq.(\ref{density}) instead of Eq.(\ref{WF}) when $H(R)>0$. The second term in Eq.(\ref{density1}) is related to the coherence of the quantum state of $S$. When $H(R)=0$, it turns out both density matrices are mathematically equivalent, as shown in the following theorem. 
\begin{theorem} 
If the entanglement measure $H(R)=0$, $\hat{\rho}_S^{\prime} = \hat{\rho}_S$.
\end{theorem}
Proof is left to Section \ref{subsect:theorem2}. Essentially, when $H(R)=0$, the coherence term in Eq.(\ref{density1}) is equal to $\hat{\rho}_S$ multiplied by a constant, effectively making $\hat{\rho}_S^{\prime}$ and $\hat{\rho}_S$ differ only by a constant. 

We see that there are three mathematical tools to describe a quantum system, namely, the relational matrix $R$, the reduced density matrix $\rho_S$, and the wave function $|\psi\rangle_{S}$. They are equivalent in terms of calculating the probability of future measurement outcome. The wave function can only be used when $H(R)=0$. The reduced density matrix, on the other hand, can describe the quantum state of $S$ regardless $H(R)=0$ or $H(R)>0$. It is more generic in quantum mechanics formulation. However, in the case of $H(R)=0$, the wave function defined in Eq.(\ref{WF}) reflects better the physical meaning of a superposition quantum state. Both $\rho_S$ and $|\psi\rangle_{S}$ are derived from $R$. This confirms that $R$ is a more fundamental variable in quantum mechanics formulation.

In deriving Eq.(\ref{WF}), we assume observer $\cal{O}_E$ who is local to apparatus $A'$ has the latest information of the relational matrix $R$ and using the same reference frame $F$. $\cal{O}_E$ then comes to an equivalent description of $S$ using the reduced density matrix, as shown in Eq.(\ref{indirectProb}). This is significant since it gives the meaning of objectivity of a quantum state. Objectivity can be defined as the ability of different observers coming to a consensus independently~\cite{Zurek03}. On the other hand, if $\cal{O}_E$ is out of synchronization on the latest information of $R$, for instance, there is update on $R$ due to measurement and not known to $\cal{O}_E$, $\cal{O}_E$ can have different descriptions of $S$. This synchronization of latest information is operational, but it is necessary. One can argue that the quantum state is absolute to any observer, but the statement is non-operational if two observers are space-like separated, and causes the EPR paradox~\cite{Rovelli07}.

\subsection{Operator}

Although $R^{SA}_{ij}$ itself is not a probability quantity, we assume it follows some of the rules for probability calculation. For example, the multiplication rule, as seen in Eq.(\ref{probprod}). Another important rule is the summation of alternatives in the intermediate steps. Let's denote the initial relational matrix for $S$ is $R^{SA}_{init}$. Suppose there is a dynamic (either a local operation or a time evolution) that changes $S$ to a new state. The effect of the dynamics connects the initial state and new state through a matrix $R^{SS}_p$. The new relational matrix element between the $A$ and $S$ is
\begin{equation}
\label{Rpropagation}
(R^{SA}_{new})_{ij} = \sum_{k}(R^{SS}_p)_{ik}(R^{SA}_{init})_{kj}
\end{equation}
Figure \ref{fig:1} in page \pageref{fig:1} shows the meaning of Eq. (\ref{Rpropagation}). The new matrix element $(R^{SA}_{new})_{ij}$ is obtained by multiplying the initial relational matrix element $(R^{SA}_{init})_{kj}$ and the local dynamics matrix element $(R^{SS}_p)_{ik}$, then summing over all possible intermediate steps.

\begin{figure}
\begin{center}
\includegraphics[scale=0.35]{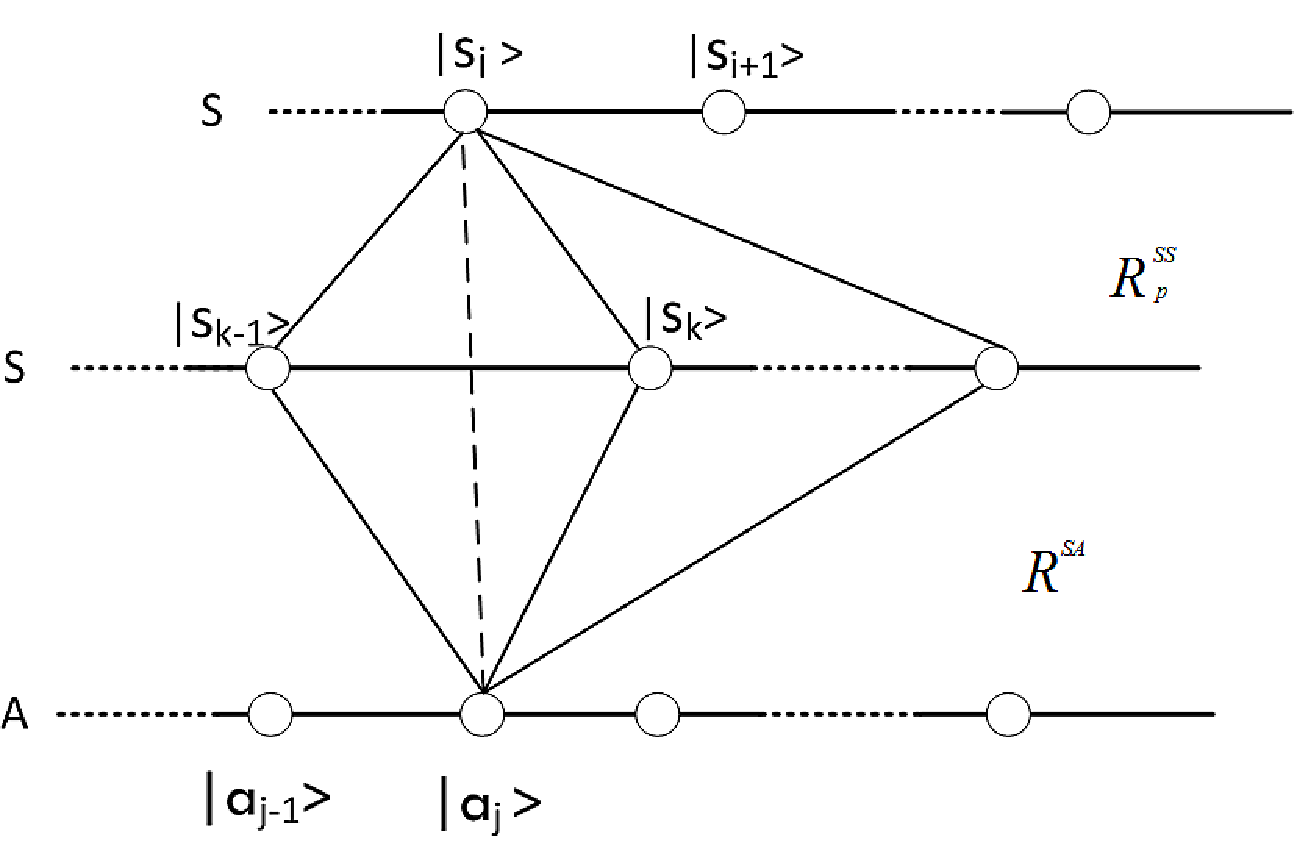}
\caption{Summation of alternatives for probability amplitude}
\label{fig:1}       
\end{center}
\end{figure}

With the notation of wave function $\varphi_i$ and reduced density matrix $\rho_S$, it is mathematically convenient to rewrite Eq.(\ref{Rpropagation}) without referring to $A$. Defined an operator $\hat{M}$ in Hilbert space ${\cal H}_S$ as $\langle s_i|\Hat{M}|s_k\rangle = (R^{SS}_p)_{ik}$, Eq.(\ref{Rpropagation}) becomes

\begin{equation}
    \label{operator1}
    (R^{SA}_{new})_{ij} = \sum_{k}M_{ik}(R^{SA}_{init})_{kj}, \quad \text{or} \quad R_{new} = MR_{init}.
\end{equation}
If $R$ is not an entangled matrix, we can sum over index $j$ in both sides of the above equation.  Referring to the definition of $\varphi_i$ we obtain $(\varphi_i)_{new} = \sum_k M_{ik}(\varphi_k)_{init}$. Substitute this into Eq.(\ref{WF}), 
\begin{equation}
    \label{operator2}
    |\psi\rangle_{new}=\hat{M}|\psi\rangle_{init}
\end{equation}
which is a familiar formulation. If $R$ is an entangled matrix, we use the reduced density formulation,
\begin{equation}
    \label{operator3}
    \rho_{new} = R_{new}(R_{new})^\dag = M\rho_{init}M^\dag.
\end{equation}
Once again, we see that change of a quantum system can be described by either the relational matrix $R$, or the reduced density matrix that traces out the information of the reference system. Both descriptions are equivalent.

\subsection{General Formulation of Time Evolution}
\label{subsec:TV}

\begin{figure}
\centering 
\includegraphics[scale=0.35]{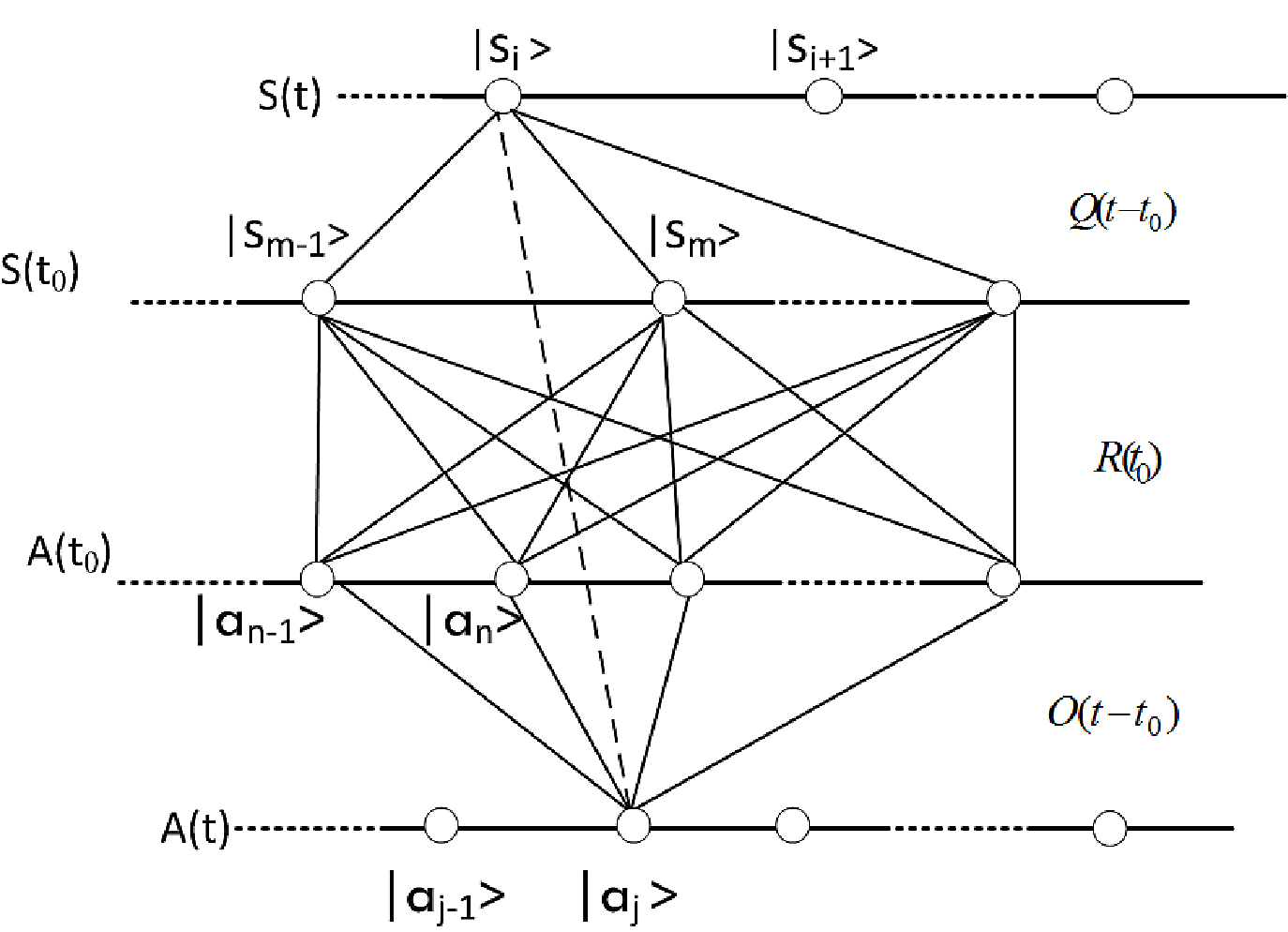}
\caption{Time evolution of probability amplitude}
\label{fig:2}       
\end{figure}

Without loss of generality, we will only consider discrete time evolution here that describes state change from initial time $t_0$ to some finite time later at $t$. By definition, there is no physical interaction between $S$ and $A$, $S$ and $A$ are evolving independently. According to Eq. (\ref{Rpropagation}), the new state that $S$ is changed to is related to the original state through a local evolution matrix $R_p^{SS}$. $R_p^{SS}$ is independent of $A$ since there is no interaction between $S$ and $A$. Similarly, the new state that $A$ is changed to is related to the original state through a local evolution matrix $R_p^{AA}$. $R_p^{AA}$ is independent of $S$. To simplify the notation, we rewrite $Q(t-t_0)=R_p^{SS}$ and $O(t-t_0)=R_p^{AA}$, and denote the initial relational matrix between $S$ and $A$ is $R(t_0)$. The time evolution of the relational matrix $R(t)$ is depicted in Figure \ref{fig:2} of page \pageref{fig:2}. Matrix element at time $t$, $R_{ij}(t)$, shown in the dot line in Figure \ref{fig:2}, is calculated by summation of all the possible intermediate steps between eigenvector $|s_i(t)\rangle$ and eigenvector $|a_j(t)\rangle$:
\begin{equation}
\label{TV}
\begin{split}
    R_{ij}^{S_tA_t}(t) & = \sum_{m,n} Q_{im}^{S_tS_0}(t-t_0) R_{mn}^{S_0A_0}(t_0) O_{nj}^{A_0A_t}(t_0-t) \\
    & = (Q(t-t_0)R(t_0)O(t_0-t))_{ij}
\end{split}
\end{equation}
The superscripts ensure the consistency of notation for the process ($S_t\rightarrow S_0 \rightarrow A_0 \rightarrow A_t$), and in the last step, they are omitted. Thus, the general formulation of the time evolution for the relational matrix is given by
\begin{equation}
\label{TVgeneral}
    R(t) = Q(t-t_0)R(t_0)O^\dag (t-t_0),
\end{equation}
where we assume the property\footnote{This property is clearer when $O$ is the representation of a unitary operator. In that case, operator $\hat{O}(t-t_0)=e^{-i\hat{H}(t-t_0)/\hbar}$ where $\hat{H}$ is a Hermitian operator. Reverting the parameter of time gives $\hat{O}(t_0-t)=e^{-i\hat{H}(t_0-t)/\hbar}=e^{i\hat{H}(t-t_0)/\hbar}= O^\dag (t-t_0)$.} $O(t_0-t)=O^\dag (t-t_0)$. For simplicity, let $t_0=0$, the reduced density matrix at time $t$ is $\rho (t)=R(t)R^\dag (t)=Q(t)R(0)O^\dag(t)O(t) R^\dag (0) Q^\dag (t)$. According to Postulate 3, in time evolution the entanglement measure is unchanged. This means the von Neumann entropy is an invariance during time evolution, $H(R(t))=H(R(0))$. One sufficient condition to meet such requirement is that $Q(t)$ and $O(t)$ are unitary matrices, consequently $\rho(t)$ and $\rho(0)$ are unitary similar matrices and have the same von Neumann entropy. However, the converse statement is not necessarily true. The condition $H(R(t))=H(R(0))$ is too weak to lead to the conclusion that $Q(t)$ and $O(t)$ are unitary matrices. We wish to find additional conditions such that $Q(t)$ and $O(t)$ are unitary\footnote{Mathematically, the Specht's Theorem and its improved version Pearcy's Theorem give the necessary and sufficient conditions for two matrices to be similar~\cite{Horn12}. This allows one to determine if $\rho(t)$ and $\rho(0)$ are unitary similar matrices. However, how such condition is related to whether $Q(t)$ and $O(t)$ are unitary matrices is not obvious. It requires further investigation.}.

\subsection{Schr\"{o}dinger Equation}
\label{subsec:TVdisent}
In the case that the initial state for $S$ and $A$ are unentangled, the eigenvalue of $\rho(0)$ has only one value $\lambda=1$ and $H(R(0))=0$. From Theorem 1, $R_{mn}(0)=c_md_n$, Eq. (\ref{TV}) becomes
\begin{equation}
\label{SE2}
\begin{split}
    R_{ij}(t) &= \sum_{m,n} Q_{im}(t) c_md_n O_{nj}^\dag(t) \\
    & = (\sum_m Q_{im}(t) c_m)( \sum_n d_nO_{nj}^\dag(t)).
\end{split}
\end{equation}
The last expression of $R_{ij}(t)$ shows it can be still decomposed into the product of two separated terms, therefore $H(R(t))=0$ as expected. By definition, the initial wave function is $\varphi_m(0)=\sum_nc_md_n=c_md_0$. At time $t$ it becomes
\begin{equation}
\label{SE3}
\begin{split}
\varphi_i(t) & = \sum_jR_{ij}(t) = \sum_m Q_{im}(t) c_m \sum_{j,n} d_nO_{nj}^\dag(t) \\
&= d(t)\sum_m Q_{im}(t) \varphi_m(0)
\end{split}
\end{equation}
where $d(t)=\sum_{jn} (d_n/d_0)O_{nj}^\dag(t)$ is a constant independent of $S$. Defined linear operator $\hat{Q}(t)$ in Hilbert space ${\cal H}_S$ as $\langle s_i|\hat{Q}(t)|s_k\rangle=Q_{ik}(t)$ and substituted Eq.(\ref{SE3}) to Eq.(\ref{WF}), the state vector
\begin{equation}
\label{SE4}
 |\psi(t)\rangle=d(t)\hat{Q}(t)|\psi(0)\rangle.
\end{equation}
Since the total probability should be preserved, $\langle\psi(t)|\psi(t)\rangle =|d|^2\langle \psi(0) |\hat{Q}^\dag\hat{Q} |\psi(0)\rangle = 1$. This is true for any initial sate $|\psi(0)\rangle$, thus, $\hat{Q}^\dag\hat{Q} = I/|d|^2$. There is an undetermined constant $|d|$.  In general, one cannot conclude that $Q(t)$ is a unitary matrix unless choosing $|d|=1$. If $|d|=1$, $d=e^{i\phi(t)}$ is an arbitrary phase. Rewritten $\hat{Q}$ as $\hat{U}$, Eq.(\ref{SE4}) becomes
\begin{equation}
\label{SE5}
 |\psi(t)\rangle=e^{i\phi(t)}\hat{U}(t)|\psi(0)\rangle = e^{i\phi(t)}e^{-i\hat{H}t/\hbar}|\psi(0)\rangle
\end{equation}
where $\hat{H}$ is a Hermitian operator for $S$. Omitting the arbitrary phase, Eq.(\ref{SE5}) is the Schr\"{o}dinger Equation. The derivation here does not give the actual expression of the Hamiltonian operator, but it manifests the fact that there is no change of entanglement measure between the observed system and the observing apparatus.

The above derivation depends on several conditions. First, there is no physical interaction between $S$ and $A$; Second, $S$ and $A$ are not entangled; Third, the total probability is preserved; Lastly, we choose $|d(t)|=1$. The first two conditions are usually what one refers as $S$ is in an isolated state. In summary, given $H(R(t))=H(R(0))$, if two more conditions are added, $H(R(0))=0$ and $|d(t)|=1$, matrix $Q(t)$ is unitary, which leads to the Schr\"{o}dinger Equation.

A special case for Eq. (\ref{TVgeneral}) to be reduced to the Schr\"{o}dinger Equation is when $O(t)=I$. With $O(t)=I$, $R(t)=U_S(t)R(0)$. Since $H(R)=0$, we can use Eq.(\ref{WF}) to calculate the wave function
\begin{equation}
\label{SE6}
\begin{split}
\varphi_i(t) &= \sum_jR_{ij}(t) = \sum_m Q_{im}(t) \sum_jR_{mj}(0) \\
& = \sum_m Q_{im}(t) \varphi_m(0)
\end{split}
\end{equation}
which is the same as Eq.(\ref{SE3}) with $d(t)=1$. The same reasoning from Eq.(\ref{SE3}) to Eq.(\ref{SE5}) is applied here. $O(t)=I$ is a very strong condition, it may not be physically sensible because any quantum system evolves as time elapses. However, this may be considered an approximation that, for a macroscopic classical apparatus, the change as a ratio to its overall state is so infinitesimal in magnitude compared to the change for a microscopic quantum state, that it can be ignored.

\subsection{Generalized Differential Equation}
\label{subsec:generalTV}
Next, we consider the more general case that $S$ and $A$ are not interacting but initially entangled, i.e., $H(R(0)) > 0$. Since entanglement measure is unchanged, $H(R(t))=H(R(0)) > 0$. $S$ and $A$ stay entangled at time $t$. $S$ is not in an isolated state. We need to describe the composite system $S+A$ as a whole relative to another unentangled apparatus $A'$. To proceed further the following theorem is introduced.
\begin{theorem}
Applying operator $\hat{Q}\otimes \hat{O}$ over the composite system $S+A$ is equivalent to change the relational matrix $R$ to $R'=QRO^T$, where the superscript $T$ represents a transposition.
\end{theorem}
The proof is left to Section \ref{subsect:theorem3}. Since the composite system $S+A$ is in isolated state, based on the result in Section \ref{subsec:TVdisent}, the overall time evolution operator $\hat{U}_{SA}$ is unitary. The state vector of the composite system at time $t$ should be $|\Psi(t)\rangle=\hat{U}_{SA}|\Psi(0)\rangle=\hat{U}_{SA}\sum_{ij}R^{SA}_{ij}|s_i\rangle|a_j\rangle$. Let $\hat{U}_{SA}=exp(-i\hat{H}_{SA}t/\hbar)$ where $\hat{H}_{SA}$ is the Hamiltonian of the composite system. Since there is no interaction between $S$ and $A$, $\hat{H}_{SA}=\hat{H}_S+\hat{H}_A$ where $\hat{H}_S$ and $\hat{H}_A$ are the Hamiltonian operators in their respective Hilbert spaces. As shown in Section \ref{subsect:decomp}, $\hat{U}_{SA}$ can be decomposed into $\hat{U}_{SA}=\hat{U}_S\otimes \hat{U}_A$. According to Theorem 3, this effectively change the relational matrix to $R(t)=U_S(t)R(0)U_A^T(t)$. Note that $U_A^T(t)$ is also a unitary matrix. This is equivalent to the general time evolution equation (\ref{TVgeneral}) with the condition that both $\hat{Q}(t)$ and $\hat{O}(t)$ are unitary.

Let's rewrite the general time evolution dynamics, Eq. (\ref{TVgeneral}), in operator notation by introducing a linear operator $\hat{R}=\sum_{ij}R_{ij}|s_i\rangle\langle a_j|$. Since $\hat{Q}(t)=\hat{U}_S(t)=exp\{-i\hat{H}_St/\hbar\}$ and $\hat{O}^\dag(t)=\hat{U}_A^T(t)=exp\{-i(\hat{H}_A^T)t/\hbar\}$, Eq. (\ref{TVgeneral}) becomes
\begin{equation}
\label{TVgeneral2}
    \hat{R}(t) = e^{-i\hat{H}_St/\hbar}\hat{R}(0)e^{-i(\hat{H}_A^T)t/\hbar}.
\end{equation}
Because $H(R)>0$, the wave function $\varphi(t)$ of $S$ cannot be defined. Consequently we cannot obtain a dynamics equation of wave function. Instead, a dynamics equation for $\hat{R}$ can be derived. Taking the derivative over $t$ of both sides of Eq.(\ref{TVgeneral2}) and noting $[exp\{i(\hat{H}_A^T)t/\hbar\}, \hat{H}_A^T]=0$, one gets
\begin{equation}
\label{TVgeneral3}
    i\hbar\frac{d\hat{R}(t)}{dt}  = \hat{H}_S\hat{R}(t) + \hat{R}(t)\hat{H}_A^T.
\end{equation}
Note that $[\hat{R},\hat{H}_A^T] \ne 0$, i.e., $\hat{R}$ and  $\hat{H}_A^T$ are non-commutative\footnote{It is easier to realize the non-commutation if using the matrix representation of Eq. (\ref{TVgeneral3}): $i\hbar (dR(t)/dt)  = H_SR(t) + R(t)H_A^T$. Since $R$ is a $N\times M$ matrix while $H_A^T$ is a $M\times M$ matrix, matrix multiplication $H_A^T\times R$ is even not possible when $N\ne M$.}. Eq. (\ref{TVgeneral3}) is a more general form of differential equation that describes the time evolution of $R$. Once solving the above equation and obtaining $R(t)$, one can calculate the probability according to Postulate 2e.

To derive a differential equation without explicitly referring to the apparatus $A$, we should use the reduced density matrix approach since $S$ and $A$ can be entangled. Given the dynamics of the relational matrix is $R(t)=U_S(t)R(0)U_A^T(t)$, the reduced density matrix of $S$ is $\rho(t) = R(t)R^\dag(t) =U_S(t)\rho(0)U_S^\dag(t)$. Defining density operator $\hat{\rho}(t)$ for $S$ such that $\langle s_i|\hat{\rho}(t)|s_j\rangle = [R(t)R^\dag (t)]_{ij}$, we can rewrite the equation to
\begin{equation}
\label{density5}
\hat{\rho}(t)  = e^{-i\hat{H}_St/\hbar}\hat{\rho}(0)e^{i\hat{H}_St/\hbar}
\end{equation}
Taking the derivative over $t$ of both sides, we obtain the Liouville$-$von Neumann equation
\begin{equation}
\label{TVgeneral5}
    i\hbar\frac{d\hat{\rho}(t)}{dt}  = \hat{H}_S\hat{\rho}(t) - \hat{\rho}(t)\hat{H}_S = [\hat{H}_S, \hat{\rho}(t)].
\end{equation}
Eqs.(\ref{TVgeneral3}) and (\ref{TVgeneral5}) give equivalent descriptions of the dynamics of quantum state of $S$. Eq.(\ref{TVgeneral5}) has the advantage of describing the time evolution of $S$ without referencing to $A$ and therefore mathematically more elegant. However, it leads to the impression that the quantum system can be described independent of the reference system. 

Eqs.(\ref{TVgeneral3}) and (\ref{TVgeneral5}) also confirm two equivalent methodologies to describe the change of quantum state of $S$: 1.) Calculate the change of relational matrix $R$ and compute the quantum probability based on Postulate 2e; 2.) Derive the wave function of the composite state for $S+A$, then trace out $A$ over the composite state to obtain $\rho_S$.

\section{Discussion and Conclusion}
\label{sec:conclusion}
\subsection{Hypotheses in the Reconstruction}
The reconstruction of quantum theory presented in this paper is based on two hypotheses. First, the relational properties between two quantum systems are more basic than the properties of one system. We take this hypothesis as a starting point to reformulate quantum mechanics. This reference system is not arbitrary. It is the apparatus, or environment, $A$, that the system $S$ has previously interacted with. Although the reference system $A$ is unique and objectively selected, it is possible that another observer does not have complete information of the interaction (or, measurement) results between $A$ and $S$. In such case she can describe $S$ differently using a different set of relational properties between $S$ and $A$. It is in this sense that we say the relational properties themselves are observer-dependent. This is indeed the main thesis of Ref.~\cite{Rovelli96}. In the example of ideal measurement described by Eq.(\ref{measurement}), supposed the measurement outcome is correspondent to eigenvector $|s_n\rangle$. For an observer that operates and reads the pointer variable of $A$, she knows the measurement outcome. At the end of the measurement, her relational description is given by $|s_n\rangle|a_n\rangle$. On the other hand, for another observer who only knows there is interaction between $S$ and $A$, but does not know the measurement outcome, the relational description is given by $\sum_i c_i|s_i\rangle|a_i\rangle$. Both descriptions are based on relational properties, and they are observer-dependent. Thus, there are two layers of relativity here. In this paper, we assume observers share the same information of the relational matrix, and focus on formulating quantum mechanics based on the relational probability amplitude. The observer-dependent aspect of the formulation is more relevant to measurement theory, which will be analyzed in an upcoming paper.

The second hypothesis is due to the realization that a real physical measurement is a bidirectional process. It is a question-and-answer, or a probe-and-response, interaction process. This bidirectional aspect of a physical measurement seems overlooked in other quantum mechanics formulations. Here we mandate that a framework of calculating probability for a potential outcome from a physical interaction must explicitly model the bidirectional process. A variable that only models unidirectional of the process cannot be considered as a real probability number because a one-way process does not model an actual measurement. Instead, such unidirectional quantity is called probability amplitude and is not necessarily a real non-negative number. The distinction of unidirectional versus bidirectional process allows us to relax the mathematical requirement on the probability amplitude and to consider it as a complex number. However, we assume it still follows some of rules in probability calculation, such as multiplication, and sum over alternatives of intermediate steps.

With these two hypotheses, a framework is developed such that the task of calculating probability in a specific measurement setup is reduced to counting the applicable measurement configurations in the joint Hilbert space for the measured system and the apparatus. It is interesting to notice that the two hypotheses philosophically echo the ideas expressed in Ref.~\cite{Smolin} that the physical world is made of processes instead of objects, and the properties are described in terms of relationships between events.

\subsection{The Apparatus System}

Although the relational properties between the quantum system $S$ and the measurement apparatus $A$, such as the probability amplitude matrix $R$, are considered as the most basic variables, there are mathematical tools that allow a quantum system $S$ to be described without explicitly calling out the apparatus system $A$. When $S$ and $A$ are unentangled, $S$ is described by a wave function that sums out the information of the reference system. When $S$ and $A$ are entangled, $S$ is described by a reduced density matrix that traces out the information of $A$. These mathematical tools enable us to develop the formulations for time evolution and measurement theory that are equivalent to those in the traditional quantum mechanics.

Except some special scenario such as that is described in the EPR argument, there is no need to explicitly call out the apparatus quantum system $A$. Mathematically it is more convenient and elegant to trace out the information of the apparatus system. However, explicitly including the apparatus system allows us to develop a framework to explain the origin of the quantum probability and to quantify the difference between time evolution and quantum measurement. 

It is interesting to notice that Ref.~\cite{Nicolaidis09} also proposes to use relational logic and category theory to deduce the basic laws of quantum mechanics. However, the formulation in Ref.~\cite{Nicolaidis09} is rather abstract. Quantum probability is introduced purely from mathematical perspective, instead of associating it with the process of actual physical measurement. How entanglement influences the probability calculation is not discussed and formulated in Ref.~\cite{Nicolaidis09}.

\subsection{Comparison with the Original RQM Theory}
The works presented here is inspired by the main idea of the original RQM theory~\cite{Rovelli96}. However, there are several significant improvements that should be pointed out. 

The works of Refs.~\cite{Rovelli96, Zurek82} establish the idea that relational properties are more basic, and a quantum system must be described relative to another quantum system. However, they do not provide a clear formulation on how a quantum system should be described relative to another system and what the basic relational properties are. On the other hand, our works gives a clear quantification of the relational property, which is the relational probability amplitude. The introduction of the relational probability amplitude is based on a detailed analysis of measurement process. It enables us to develop a framework to calculate probability during quantum measurement. We further show that the relational probability amplitude can be calculated using Feynman path integral in Section \ref{sec:PI}. 

The second improvement in this works comes from the introduction of the concept of entanglement to the RQM theory. We recognize not only that a quantum system must be described relative to another quantum system, but also that the entanglement between these two systems plays a crucial role in how the formulation the observed system is described. If there is no entanglement, the observed system can be described by a pure wave function. If there is entanglement, a reduced density matrix is more appropriate mathematical tool. In addition, entanglement measure plays a pivot role in determining a system is undergoing a time evolution or measurement process. This allows us to reconstruct both the Schr\"{o}dinger equation and the measuring theory\footnote{The reconstruction of quantum measurement theory is submitted in an upcoming paper.}. When one states that a quantum system must be described relative to another quantum system, one can further quantify this relativity via the entanglement measure between these two systems. However, the concept of entanglement is not presented in Ref~\cite{Rovelli96}. The reconstruction attempts in Ref~\cite{Rovelli96} to derive the laws of quantum mechanics based on quantum logic is rather limited since only the Schr\"{o}dinger equation is reconstructed. 

Thirdly, although a quantum system must be described relative to another quantum system, our works show that there are mathematical tools that can describe the observed system without explicitly calling out the observing system, such as the wave function and the reduced density matrix as shown in Section \ref{sec:results}. Therefore, RQM and traditional QM are compatible mathematically. This is important because it confirms that although the main idea of RQM seems radical, it does not change the practical application of quantum mechanics. Again, this point was not clear in Ref~\cite{Rovelli96}.

\subsection{Compared to the Transactional Interpretation}
\label{subsec:transaction}
The bidirectional measurement process which is important in the derivation of the measurement probability appears sharing some common ideas with the transaction model~\cite{Cramer}. In particular, the transaction model requires a handshake between a retarded ``offered wave" from an emitter and an advanced ``confirmation wave" from an absorber to complete a transaction in a quantum event. This is a bidirectional process. While it is encouraging to note that the bidirectional nature of a quantum event has been recognized in the transaction model, there are several fundamental differences between the transaction model and the bidirectional measurement framework presented here. First, the transaction model considers the offered wave and the confirmation wave as real physical waves. In our framework, we do not assume such waves existing. Instead, the probing and responding are just two aspects in a measurement event, and we require the probability calculation to faithfully model such bidirectional process. Second, in the transaction model, the randomness of measurement outcomes is due to the existence of different potential future absorbers. Thus, the randomness in quantum mechanics depends on the existence of absorbers. There is no such assumption in our framework. Third, the transaction model derives that the amplitude of the confirmation wave at the emitter is proportional to the modulus square of the amplitude of the offer wave, which is related to the probability of completing a transaction with the absorber. This appears to be ambiguous since it suggests the confirmation wave is also a probability wave. In our formulation, we only focus on how the measurement probability is calculated, and clearly point out that the wave function is just a mathematical tool.

\subsection{Limitations and Outlook}
\label{sec:outlook}
The framework to calculate the measurement probability in Section \ref{subsec:postulates} is the key to our reformulation. However, it is essentially based on an operational model from a detailed analysis of bidirectional measurement process, instead of being derived from more fundamental physical principles. In particular, there may be better justification for Eq. (\ref{conjugate}). The current model is only served as a step to deepen our understanding of relational quantum mechanics. It is desirable to continue searching for more fundamental physical principles to justify the calculation of measurement probability. The fact the $R^{SA}_{ij}$ is a complex number means that this variable actually comprises two independent variables, the modulus and the phase. This implies that more degrees of freedom are needed to have a complete description of a quantum event. Stochastic mechanics, for instance, introduces forward and backward velocities instead of just one classical velocity to describe the stochastic dynamics of a point particle. With the additional degree of freedom, two stochastic differential equations for the two velocities are derived. Then, through a mathematical transformation of two velocity variables in $\mathbb{R}$ into one complex variable in $\mathbb{C}$, the two differential equations turn into the Schr\"{o}dinger equation~\cite{Nelson, Nelsonbook, Yasue, Guerra, Yang2021}. It will be interesting to investigate if $R^{SA}_{ij}$ can be implemented in the context of stochastic mechanics, where we expect $R^{SA}_{ij}$ will be decomposed to two independent variables in $\mathbb{R}$ and each of them is a function of velocity variables.

As discussed in the introduction section, the RQM principle consists two aspects. First, we need to reformulate quantum mechanics relative to a QRF which can be in a superposition quantum state, and show how quantum theory is transformed when switching QRFs. In this aspect, we accept the basic quantum theory as it is, including Schr\"{o}dinger equation, Born's rule, von Neunman measurement theory, but add the QRF into the formulations and derive the transformation theory when switching QRFs~\cite{Brukner, Hoehn2018, Yang2020}. Second, we go deeper to reformulate the basic theory itself from \textit{relational properties}, but relative to a fixed QRF. Here the fixed QRF is assume to be in a simple eigen state. This is what we do in the present work. A complete RQM theory should combine these two aspects together. This means one will need to reformulate the basic quantum theory from relational properties and relative to a quantum reference frame that exhibits superposition behavior. Therefore, a future step is to investigate how the relational probability amplitude matrix should be formulated when the QRF possesses superposition properties, and how the relational probability amplitude matrix is transformed when switching QRFs.

There are other limitations that are worth to mention here. The formulation assumes a finite Hilbert space for either the observed quantum system or the observing apparatus. It is desirable to extend the formulation for Hilbert space with infinite dimension. It is mathematically more cumbersome to calculate the wave function from a relational matrix than to just assume an observer independent wave function. Mathematical rigorousness is needed for some of the derivations. For instance, given the general time evolution dynamics in Eq. (\ref{TVgeneral}), it is left unanswered on what the sufficient and necessary condition should be in order to keep the entanglement measure as an invariance. Section \ref{subsec:TVdisent} only gives several sufficient conditions that lead to the Schr\"{o}dinger Equation. Furthermore, implementing the relational properties between $S$ and $A$ with one definite matrix implies that the composite system $S+A$ is in a pure state. $S+A$ can be in a mixed state and described by an ensemble of relational matrices. A rigorous mathematical treatment for mixed state is desirable, especially when $\rho_{SA}$ is not a separable mixed state. It should be also noted that only non-relativistic quantum mechanics is considered here.

\subsection{Conclusions}
We have shown that quantum mechanics can be constructed by shifting the starting point from the independent properties of a quantum system to the relational properties among quantum systems. This idea, combined with the emphasis that a physical measurement is a bidirectional interaction process, enables us to propose a framework to calculate the probability of outcome when measuring a quantum system. Quantum probability is proportional to the summation of weights representing the bidirectional measurement process from all applicable configuration in the joint Hilbert space of the measured and measuring composite system. This postulate explains why the quantum probability is the absolute square of a complex number when there is no entanglement. The wave function of the observed system is simply the summation of relational probability amplitudes. If there is entanglement between the measured and measuring composite system, the way probability is calculated is adjusted due to the presence of correlation. In essence, quantum mechanics demands a new set of rules to calculate probability of a potential outcome from a physical interaction in the quantum world. Quantum theory does not describe directly measurable physical properties such as force, length, etc. Instead it deals with quantity such as probability amplitude, and provides a set of rules to connect to those measurable physical properties. In this sense, quantum mechanics is a probability theory for describing the process of measuring a quantum system through interaction. 

Based on the postulates, formulations for time evolution and quantum measurement can be reconstructed. Schr\"{o}dinger Equation is derived when the observing system is in an isolated state. Although the theory developed in this work is mathematically equivalent to the traditional quantum mechanics, there are several significant implications of this formulation. First, the reformulation shows that relational property can be the most fundamental element to construct quantum mechanics. Second, it brings new insight on the origin of the quantum probability. Third, path integral formulation is generalized to formulate the reduced density matrix of a quantum system. This may pave the way to extend the reformulation to quantum field theory and deserves further research. Finally, as with other efforts of reformulating quantum mechanics, it is always interesting to recognize a new perception on a traditional theory. The hope is that the reformulation presented here can be one step towards a better understanding of quantum mechanics.

\section{Methods}

\subsection{Proof of Eq.(\ref{generalizedProb})}
\label{subsec:generalizedProb}
To prove Eq.(\ref{generalizedProb}), we perform a transformation of eigenbasis. The initial eigenbasis for $S$ is $\{ |s_i\rangle \}$ and the relational matrix is $R$. If we introduce another set of eigenbasis $\{ |s^\prime_i\rangle \}$ such that the first eigenvector is $|s^\prime_0\rangle$ is $|\chi\rangle$. Denote the unitary matrix that relates the two sets of eigenbasis as $U$. We have $U|\chi\rangle = |s^\prime_0\rangle$, or, 
\begin{equation}
    |\chi\rangle = U^\dag|s^\prime_0\rangle
\end{equation}
From the definition of wave function, we have $|\chi\rangle = \{\phi_0, \phi_1, \ldots, \phi_N \}^T$, where $\phi_i = \sum_jQ_{ij}$. Substitute this into the above equation, we get $U^\dag_{i0} = \phi_i$, i.e., 
\begin{equation}
\label{A2}
    U_{0i}=\sum_jQ^*_{ij}
\end{equation}
In the new eigenbasis, the original relational matrix $R$ is transformed to $R^\prime = UR$. The probability of finding $S$ described by state vector $\chi$ is correspondent to the probability to find $S$ in engeinvector $|s^\prime_0\rangle$, which according to Eq.(\ref{probability2}) is
\begin{equation}
\begin{split}
    p(\chi|\psi) & = |\sum_jR^\prime_{0j}|^2 = |\sum_j(UR)_{0j}|^2\\
    & = |\sum_j\sum_iU_{0i}R_{ij}|^2  = |\sum_{ij}\sum_kQ^*_{ik}R_{ij}|^2\\
    &= |\sum_{jk}(\sum_iQ^\dag_{ki}R_{ij})|^2  = |\sum_{jk}(Q^\dag R)_{kj}|^2 \\
    & = |\sum_i (\sum_kQ^*_{ik})(\sum_jR_{ij})|^2 \\
     & = |\sum_i \phi^*_i\psi_i|^2 = \| \langle\chi | \psi\rangle \|. 
\end{split}
\end{equation}
In the first step of the second line, we use the relation Eq.(\ref{A2}).

\subsection{Proof of Theorem 1}
\label{subsect:theorem1}
According to the singular value decomposition, the relational matrix $R$ can be decomposed to $R = UDV$, where $D$ is rectangular diagonal and both $U$ and $V$ are $N\times N$ and $M\times M$ unitary matrix, respectively. This gives $\rho = RR^\dag = U(DD^\dag) U^\dag$. If $H(R)=0$, matrix $\rho$ is a rank one matrix, therefore $DD^\dag$ is $diag\{1,0,0...\}$. This means $D$ is a rectangular diagonal matrix with with only one eigenvalue $e^{i\phi}$. Expanding the matrix product $R=UDV$ gives 
\begin{equation}
    R_{ij}=\sum_{nm}U_{in}D_{nm}V_{mj}=U_{i1}e^{i\phi}V_{1j}.
\end{equation}
We just choose $c_i=U_{i1}$ and $d_j=e^{i\phi}V_{1j}$ to get $R_{ij}=c_i d_j$. Conversely, if $R_{ij}=c_i d_j$, $R$ can be written as outer product of two vectors,
\begin{equation}
R = \begin{pmatrix} c_1 & c_2 & \ldots  & c_n \end{pmatrix}^T \times \begin{pmatrix} d_1 & d_2 & \ldots & d_m \end{pmatrix}.
\end{equation}
Considering vector $C_1=\{c_1, c_2, \ldots,c_n\}$ as an eigenvector in Hilbert space ${\cal H}_S$, one can use the Gram-Schmidt procedure~\cite{Nelson00} to find orthogonal basis set $C_2, \ldots, C_n$. Similarly, considering vector $D_1=\{d_1, d_2, \ldots,d_m\}$ as an eigenvector in Hilbert space ${\cal H}_A$, one can find orthogonal basis set $D_2, \ldots, D_m$. Under the new orthogonal eigenbasis, $R$ becomes a rectangular diagonal matrix $D=diag\{1,0,0...\}$. Therefore $R=UDV$ where $U$ and $V$ are two unitary matrices associated with the eigenbasis transformations. Then $\rho = RR^\dag=U(DD^\dag) U^\dag$, and $DD^\dag=diag\{1,0,0...\}$ is a square diagonal matrix. Since the eigenvalues of similar matrices are the same, the eigenvalues of $\rho$ are (1, 0, ...), thus $H(R)=0$. 

\subsection{Proof of Theorem 2}
\label{subsect:theorem2}
Assuming $S+A$ is in a pure state, we use the Von Neumann entropy H(R) as entanglement measure. If $H(R)=0$, by virtue of Theorem 1, $R_{ij}=c_id_j$. Assuming both $\cal{O}_I$ and $\cal{O}_E$ share the same knowledge of $R_{ij}$. The reduced density matrices relative to each observer are calculated as
\begin{equation}
\begin{split}
    \hat{\rho}_S^{\prime}  & = \sum_{ii'}(\sum_{jj'}R_{ij}R^*_{i'j'})|i\rangle\langle i'|\\
    &= \sum_{ii'}c_i(c_{i'})^*(\sum_{jj'}d_jd_{j'})|i\rangle\langle i'|\\
    & =d_A\sum_{ii'}c_i(c_{i'})^*|i\rangle\langle i'| \\
    \hat{\rho}_S & = \sum_{ii'}(\sum_kR_{ik}R^*_{i'k})|i\rangle\langle i'|\\
    &=\sum_{ii'}c_i(c_{i'})^*(\sum_k|d_k|^2)|i\rangle\langle i'| \\
    &= d_{A'}\sum_{ii'}c_i(c_{i'})^*|i\rangle\langle i'|
\end{split}
\end{equation}
where $d_A$ and $d_{A'}$ are two constant. $\hat{\rho}_S$ and $\hat{\rho}_S^{\prime}$ only differ by a constant when $H(R)=0$. Since $Tr(\hat{\rho}_S^{\prime})=Tr(\hat{\rho}_S)=1$, we can simply choose $d_A=d_{A'}$ so that $\hat{\rho}_S=\hat{\rho}_S^{\prime}$.  

\subsection{Proof of Theorem 3}
\label{subsect:theorem3}
Denote the initial state vector of the composite system as $|\Psi_0\rangle=\sum_{ij}R_{ij}|s_i\rangle|a_j\rangle$. Apply the composite operator $\hat{Q}(t)\otimes\hat{O}(t)$ to the initial state,
\begin{equation}
\begin{split}
 |\Psi_1\rangle & = (\hat{Q}\otimes \hat{O}) \sum_{ij} R_{ij} |s_i\rangle \otimes |a_j\rangle \\
 & = \sum_{ij}R_{ij}\hat{Q}|s_i\rangle\otimes \hat{O}|a_j\rangle \\
 & = \sum_{ij}\sum_{mn}R_{ij}Q_{mi}O_{nj}|s_m\rangle\otimes |a_n\rangle \\
 & = \sum_{mn}(\sum_{ij}Q_{mi}R_{ij}O^T_{jn})|s_m\rangle\otimes |a_n\rangle \\
 & =\sum_{mn}(QRO^T)_{mn}|s_m\rangle|a_n\rangle
\end{split}
\end{equation}
where $T$ represents the transposition of matrix. Compared the above equation to Eq.(\ref{WF}) for the definition of $|\Psi_1\rangle$, it is clear that the relational matrix is changed to $R'=QRO^T$.  

\subsection{Decomposition of the Unitary Operator of a Bipartite System}
\label{subsect:decomp}
Here we show that if there is no interaction between $S$ and $A$, a global unitary operator for the composite system $S+A$ is decomposed into the tensor product of two local unitary operators. Let $\{|s_i\rangle\}$ be the orthogonal eigenbasis of $\hat{H}_S$, $\hat{H}_S|s_i\rangle=E_i^S|s_i\rangle$. Recall that the definition of a function of operator $\hat{H}$ is
\begin{equation}
    f(\hat{H}) = \sum_i f(E_i)|s_i\rangle\langle s_i|
\end{equation}
Based on this definition, $\hat{U}_S = exp\{-(i/\hbar)\hat{H}_St\} = exp\{-(i/\hbar)E_i^St\}|s_i\rangle\langle s_i|$. Similarly, let $\{|a_j\rangle\}$ be the orthogonal eigenbasis of $\hat{H}_A$, $\hat{H}_A|a_j\rangle =E_j^A|a_j\rangle$ and $\hat{U}_A = exp\{-(i/\hbar)E_j^At\}|a_j\rangle\langle a_j|$. When there is no interaction between $S$ and $A$, $\hat{H}_{SA}=\hat{H}_S+\hat{H}_A$ where $\hat{H}_S$ and $\hat{H}_A$ are the Hamiltonian operators in their respective Hilbert spaces, thus $\hat{U}_{SA}= exp\{-(i/\hbar)(\hat{H}_S+\hat{H}_A)t)\}$. According to Postulate 4, the set $\{|s_i\rangle|a_j\rangle\}$ forms the orthogonal eigenbasis for $\hat{H}_{SA}$, so that $\hat{H}_{SA}|s_i\rangle|a_j\rangle=(E_i^S+E_j^A)|s_i\rangle|a_j\rangle$ and $exp\{-(i/\hbar)(\hat{H}_S+\hat{H}_A)t)|s_i\rangle|a_j\rangle = exp\{-(i/\hbar)(E_i^S+E_j^A)t)|s_i\rangle|a_j\rangle$. From the definition of operator function,
\begin{equation}
\begin{split}
    \hat{U}_{SA} & = \sum_{ij}f(E_{ij})|s_i\rangle|a_j\rangle\langle s_i|\langle a_j| \\ &=\sum_{ij}exp\{-(i/\hbar)(E_i^S+E_j^A)t)|s_i\rangle|a_j\rangle\langle s_i|\langle a_j| \\
    &= \sum_i exp\{-(i/\hbar)E_i^St\}|s_i\rangle\langle s_i| \\
    & \otimes \sum_j exp\{-(i/\hbar)E_j^At\}|a_j\rangle\langle a_j| \\
    & = \hat{U}_S \otimes \hat{U}_A. 
\end{split}
\end{equation} 

\subsection{Path Integral Implementation}
\label{sec:PI}
This section briefly describes how the relational probability amplitude can be calculated using the Path Integral formulation. Without loss of generality, the following discussion just focuses on one dimensional space-time quantum system. In the Path Integral formulation, the probability to find a quantum system moving from a point $x_a$ at time $t_a$ to a point $x_b$ at time $t_b$ is the absolute square of a probability amplitude, i.e., $P(b, a)=|K(b, a)|^2$. The probability amplitude is postulated as the sum of the contribution of phase from each path~\cite{Feynman05}:
\begin{equation}
\label{PIWF}
K(b, a)=\frac{1}{N}\sum_{path}e^{(i/\hbar)S_p(x(t))}
\end{equation}
where $N$ is a normalization constant, and $S_p (x(t))$ is the action along a particular path from point $x_a$ to point $x_b$. The action is defined as $S_p (x(t))=\int_{t_a}^{t_b}L(\dot{x}, x, t)dt$ where $L$ is the Lagrangian of the system. Since there is infinite number of possible paths from point $x_a$ to point $x_b$, more precisely the summation in Eq.({\ref{PIWF}) should be replaced by an integral
\begin{equation}
\label{PIWF2}
K(b, a)=\int_{a}^{b}e^{(i/\hbar)S_p(x(t))}{\cal{D}}x(t)
\end{equation}
where ${\cal{D}}x(t)$ denotes integral over all possible paths from point $x_a$ to point $x_b$. It is the wave function for $S$ moving from $x_a$ to $x_b$~\cite{Feynman05}. The wave function of the particle at position $x_b$ is
\begin{equation}
    \label{PIWF3}
    \varphi(x_b, t_b)=\int_{-\infty}^{\infty}K(x_b, t_b; x_a, t_a)\varphi(x_a, t_a)dx_a
\end{equation}
where $\varphi(x_a, t_a)$ is the wave function of the particle at position $x_a$. Eq.(\ref{PIWF3}) is the integral form of the Schr\"{o}dinger Equation (\ref{SE5}).

Now let's consider how the relational matrix element can be formulated. At a particular time $t_a$, we denote the matrix element as $R(x_a; y_a)$. Here the coordinates $x_a$ and $y_a$ act as indices to the system $S$ and apparatus $A$, respectively. From time $t_a$ to $t_b$, suppose $S$ moves from $x_a$ to $x_b$, and $A$ moves from $y_a$ to $y_b$, the relational matrix element is written as $R(x_b, x_a; y_b, y_a)$. Borrowing the ideas described in Eq.(\ref{PIWF2}), we propose that
\begin{equation}
    \label{PIR}
    \begin{split}
    R(x_b, x_a; y_b, y_a) &= \int_a^b \int_{a}^{b}e^{(i/\hbar)S^{SA}_p(x(t), y(t))}\\
    &\times {\cal{D}}x(t){\cal{D}}y(t)
    \end{split}
\end{equation}
where the action $S_p^{SA}(x(t), y(t))$ consists three terms
\begin{equation}
    \label{actions}
    \begin{split}
    S^{SA}_p(x(t), y(t)) &= S^S_p(x(t)) + S^A_p(y(t)) \\
    & + S^{SA}_{int}(x(t), y(t)).
    \end{split}
\end{equation}
The last term is the action due to the interaction between $S$ and $A$ when each system moves along its particular path. Eq.(\ref{PIR}) is considered an extension of Postulate 1. We can validate Eq.(\ref{PIR}) by deriving formulation that is consistent with traditional path integral. Suppose there is no interaction between $S$ and $A$. The third term in Eq.(\ref{actions}) vanishes. Eq.(\ref{PIR}) is decomposed to product of two independent terms,
\begin{equation}
    \label{PIR2}
    \begin{split}
    R(x_b, x_a; y_b, y_a) & = \int_a^b e^{(i/\hbar)S^{S}_p(x(t)) }{\cal{D}}x(t)\\
    & \times \int_a^b e^{(i/\hbar)S^{A}_p(y(t)) }{\cal{D}}y(t)
    \end{split}
\end{equation}
Noticed that the coordinates $y_a$ and $y_b$ are equivalent of the index $j$ in Eq.(\ref{WF}), the wave function of $S$ can be obtained by integrating $y_a$ and $y_b$ over Eq.(\ref{PIR2})
\begin{equation}
    \label{PIWF4}
    \begin{split}
    \varphi(x_b, x_a) & = \int\int^{\infty}_{-\infty}R(x_b, x_a; y_b, y_a)dy_ady_b \\
    &=\{\int_a^b e^{(i/\hbar)S^{S}_p(x(t)) }{\cal{D}}x(t)\}\\
    &\times \{\int\int^{\infty}_{-\infty}\int_a^b e^{(i/\hbar)S^{A}_p(y(t)) }{\cal{D}}y(t)dy_ady_b\} \\
    &=c\int_a^b e^{(i/\hbar)S^{S}_p(x(t)) }{\cal{D}}x(t)
    \end{split}
\end{equation}
where constant $c$ is the integration result of the second term in step two. The result is the same as Eq.(\ref{PIWF2}) except an unimportant constant.

Next, we consider the situation that there is entanglement between $S$ and $A$ as a result of interaction. The third term in Eq.(\ref{actions}) does not vanish. We can no longer define a wave function for $S$. Instead, a reduced density matrix should be used to describe the state of the particle, $\rho = RR^\dag$. From Eq.(\ref{PIR}), the element of the reduced density matrix is 
\begin{equation}
    \label{PIRho}
    \begin{split}
    \rho(x_b, x'_b; x_a, x'_a) &= \sum_ {y_a,y_b}\int_{x_a}^{x_b}\int_{x'_a}^{x'_b}\int_{y_a}^{y_b}\int_{y_a}^{y_b}e^{(i/\hbar)\Delta S} \\
    &\times {\cal{D}}x(t){\cal{D}}x'(t){\cal{D}}y(t){\cal{D}}y'(t)\\
    \text{where}\quad \Delta S &= S^S_p(x(t)) - S^S_p(x'(t)) \\ 
    & + S^A_p(y(t))  - S^A_p(y'(t)) \\
    & + S^{SA}_{int}(x(t), y(t)) \\
    & - S^{SA}_{int}(x'(t), y'(t)).
    \end{split}
\end{equation}
The path integral over ${\cal{D}}y'(t)$ takes the same end points $y_a$ and $y_b$ as the path integral over ${\cal{D}}y(t)$. After the path integral, a summation over $y_a$ and $y_b$ is performed. Eq.(\ref{PIRho}) is equivalent to the $J$ function introduced in Ref~\cite{Feynman05-2}. We can rewrite the expression of $\rho$ using the \textit{influence functional}, $F(x(t), x'(t))$,
\begin{equation}
    \label{PIIF}
    \begin{split}
    \varrho(x_b, x'_b; x_a, x'_a) &= \frac{1}{Z} \int_{x_a}^{x_b}\int_{x'_a}^{x'_b}e^{(i/\hbar)[S^S_p(x(t)) - S^S_p(x'(t))]} \\
    &\times F(x(t), x'(t)){\cal{D}}x(t){\cal{D}}x'(t) \\
    F(x(t), x'(t)) & = \sum_ {y_a,y_b}\int_{y_a}^{y_b}\int_{y_a}^{y_b}e^{(i/\hbar)\Delta S'}\\
    &\times {\cal{D}}y(t){\cal{D}}y'(t)\\
    \text{where}\quad \Delta S' &= S^A_p(y(t))  - S^A_p(y'(t)) \\
    & + S^{SA}_{int}(x(t), y(t)) \\
    & - S^{SA}_{int}(x'(t), y'(t)).
    \end{split}
\end{equation}
Where $Z=Tr(\rho)$ is a normalization factor to ensure $Tr(\varrho) = 1$. The reduced density matrix allows us to calculate the probability of the system changing from one state to another, for instance, the probability of the system initially in a state $\chi(x_a)$ transitioning to another state $\psi(x_b)$. This is similar to calculate the probability of an ideal measurement that specifies the initial state is $\chi(x_a)$ and the final state is $\psi(x_b)$. Defining a project operator $\Hat{P}=|\chi(x_a)\psi(x_b)\rangle\langle \chi(x_a)\psi(x_b)|$, the probability is calculated, similar to Eq.(\ref{indirectProb}), as 
\begin{equation}
    \label{PIprob}
    \begin{split}
    p(\chi,\psi) &=Tr(\varrho\Hat{P}) \\
    &=\int\int\int\int \psi^*(x'_b)\psi(x_b)\varrho(x_b, x'_b; x_a, x'_a) \\
    & \times \chi(x_a)\chi^*(x'_a)dx_adx_bdx'_adx'_b
    \end{split}
\end{equation}
This is equivalent to the result in Ref~\cite{Feynman05-3}. To find the particle moving from a particular position $\bar{x}_a$ at time $t_a$ to another particular position $\bar{x}_b$ at time $t_b$, we substitute $\chi(x_a)=\delta(x_a-\bar{x}_a)$ and $\chi(x_b)=\delta(x_b-\bar{x}_b)$ into Eq.(\ref{PIprob}),
\begin{equation}
    \label{PIdensityProb}
    \begin{split}
    p(\bar{x}_b, \bar{x}_a) &= \int\int\int\int\varrho(x_b, x'_b; x_a, x'_a)\delta (x_b - \bar{x}_b) \\
    &\times \delta (x'_b - \bar{x}_b) \delta (x_a - \bar{x}_a) \delta (x'_a - \bar{x}_a)\\
    &\times dx_b dx'_b dx_a dx'_a \\
    & = \varrho(\bar{x}_b, \bar{x}_b; \bar{x}_a, \bar{x}_a).
    \end{split}
\end{equation}

In summary, we show that the relational probability amplitude introduced in Postulate 1 can be explicitly calculated through Eq.(\ref{PIR}). With this definition and the results in Section \ref{sec:results}, we obtain the formulations for wave function in Eq. (\ref{PIWF4}) and probability in Eq.(\ref{PIprob}) that are the consistent with those in traditional path integral formulation. The reduced density expression in Eq.(\ref{PIRho}), although equivalent to the $J$ function in Ref~\cite{Feynman05}, has richer physical meaning. For instance, we can calculate the entanglement measure from the reduced density matrix.  

\newpage




\section{Author Information}
\textbf{Affiliations} \\
Qualcomm, 5775 Morehouse Drive, San Diego, CA 92121, USA\\
J. M. Yang 

\textbf{Contributions}\\
J.M.Y. designed the study, conceived the ideas, performed the mathematical calculation, and wrote the manuscript.

\textbf{Competing interests}\\
The author declares no competing interests as defined by Nature Research, or other interests that might be perceived to influence the results and/or discussion reported in this paper.

\end{document}